\documentclass[useAMS,usenatbib]{mn2e}
\usepackage{graphics}
\usepackage{epsfig}
\usepackage{natbib}

\title[Local Properties of WMAP Cold Spot]{Local Properties of WMAP Cold Spot}
\author[Wen Zhao]{Wen Zhao \\
Key Laboratory for Researches in Galaxies and Cosmology, University of Science and Technology of China, Hefei, 230026, China\\
Department of Astronomy, University of Science and Technology of China, Hefei, 230026, China
\\ Niels Bohr Institute and DISCOVERY Center, Blegdamsvej 17, 2100 Copenhagen, {\O},  Denmark}
\begin{document}

\date{Accepted ........ Received .......; in original form .......}

\pagerange{\pageref{firstpage}--\pageref{lastpage}} \pubyear{2012}

\maketitle

\label{firstpage}

\begin{abstract}
We investigate the local properties of WMAP Cold Spot (CS) by defining the local statistics: mean temperature,
variance, skewness and kurtosis. We find that, compared with the \emph{coldest spots}
in random Gaussian simulations, WMAP CS deviates from
Gaussianity at $\sim 99\%$ significant level. In the meanwhile,
when compared with the spots at the same position in the
simulated maps, the values of local variance and
skewness around CS are all systematically larger in the scale of
$R>5^{\circ}$, which implies that WMAP CS is a large-scale
non-Gaussian structure, rather than a combination of some small
structures. This is consistent with the finding that the non-Gaussianity
of CS is totally encoded in the WMAP low multipoles.
Furthermore, we find that the cosmic texture can excellently
explain all the anomalies in these statistics.
\end{abstract}


\section{Introduction}
Cosmic Microwave Background (CMB) radiation is one of the most
ancient fossils of the Universe. The observations of the NASA
Wilkinson Microwave Anisotropy Probe (WMAP) satellite on the CMB
temperature and polarization anisotropies have put tight
constraints on the cosmological parameters \citep{komatsu2011}. In
addition, some anomalies in CMB field have also been reported soon
after the release of the WMAP data (see \citep{bennett2011} as a
review). Among these, an extremely Cold Spot (CS) centered at Galactic coordinate ($l=209^{\circ}$, $b=-57^{\circ}$) with a characteristic
scale about $10^{\circ}$ was detected
in the Spherical Mexican Hat Wavelet (SMHW) non-Gaussian analyses
\citep{vielva2004}.

Compared with the distribution derived from the isotropic and
Gaussian CMB simulations, due to this CS, the SMHW coefficients of
WMAP data have an excess of kurtosis \citep{cruz2005}. In
addition, several non-Gaussian statistics, such as the amplitude
and area of the cold spot, the higher criticism and so on, have
also been applied to identify this WMAP CS
\citep{cruz2007a,cruz2005,cayon2005,naselsky2010,zhang2010,vielva2010}.
Since then, various alternative explanations for the CS have been
proposed, including the possible foregrounds
\citep{cruz2006,hansen2012}, Sunyaev-Zeldovich effect
\citep{cruz2008}, the supervoid in the Universe
\citep{inoue2006,inoue2007,inoue2012}, and the cosmic texture
\citep{cruz2007b,cruz2008}. In order to distinguish different
interpretations, some analyses have been carried out, such as the
non-Gaussian tests for the different detectors and different
frequency channels of WMAP satellite \citep{vielva2004,cruz2005},
the investigation of the NVSS sources
\citep{rudnick2007,smith2010}, the survey around the CS with
MegaCam on the Canada-France-Hawaii Telescope \citep{granett2009},
the redshift survey using VIMOS on VLT towards CS
\citep{bremer2010}, and the cross-correlation between WMAP and
Faraday depth rotation map \citep{hansen2012}.

Nearly all the interpretations of CS are related to the local
characters of the CMB field, so the studies on the local properties
of CS are necessary. In this paper, we shall propose a set of
novel non-Gaussian statistics, i.e. the local mean temperature,
variance, skewness and kurtosis, to study the local properties of
the CMB field. By altering the radium of the cap around CS, we
study the local properties of CS at different scales. Compared
with the \emph{coldest spots} in the random Gaussian simulations, we find
the local non-Gaussianity of WMAP CS, i.e. it deviates from
Gaussianity at $\sim 99\%$ significant level. Furthermore, we find
the significant difference between WMAP CS and Gaussian
simulations at all the scales $1^{\circ}\le R\le15^{\circ}$.

To study the possible origin of WMAP CS, we have also compared it
with the spots at the same position of the simulated Gaussian samples. We
find that different from the general properties of the
foregrounds, the point sources or various local contaminations, in
the small scales the local variance, skewness and kurtosis values
of CS are not significantly large, except for its coldness in
temperature. However, after the careful comparison with Gaussian
simulations, we find that when $R>5^{\circ}$ the local variance
and skewness are systematically large. This implies that CS
prefers a large-scale non-Gaussian structure. In order to confirm
it, we repeat the analyses adopted by many authors, where the
statistics of temperature and kurtosis in SMHW domain are used. We
apply these analyses to the WMAP data with different $l_{\max}$,
and find that nearly all the non-Gaussianities of CS are encoded
in the low multipoles $l\le40$.

It was claimed that the cosmic texture seemed to be the most promising
explanation \citep{cruz2007b,cruz2008}, by
investigating the temperature and area of CS. In order to check
this explanation by our local statistics, we superimpose a similar
cosmic texture into the simulated Gaussian samples, and calculate the
local statistics of the CMB fields. We find that the excesses of the
local statistics of WMAP CS can be excellently explained by this
non-Gaussian structure. So our local analyses of the CS supports
the cosmic texture explanation.

The rest of the paper is organized as follows: In Section 2, we
introduce the WMAP data, which will be used in the analyses. In
Section 3, we define the local statistics and apply them to WMAP
data. In Section 4, the dependence of the WMAP non-Gaussianities on the value of $l_{\max}$ are studied, which shows that the
non-Gaussian signals are all encoded in the low multipoles.
Section 5 summarizes the main results of this paper.

\section{WMAP data and simulations}
In our analyses, we shall use the WMAP data including the VW7 map,
ILC7 map and NILC5 map.

\subsection{VW7}
The CMB temperature maps derived from the WMAP observations are
pixelized in HEALPix format with the total number of pixels
$N_{\rm pix}=12N^2_{\rm side}$. In our analyses, we use the 7-year
WMAP data for V and W frequency bands with $N_{\rm side}=512$. The
linearly co-added map (written as ``VW7") is constructed by using an
inverse weight of the pixel-noise variance
$\sigma_0^2/\bar{N}_{\rm obs}$, where $\sigma_0$ denotes the pixel
noise for each differential assembly (DA) and $\bar{N}_{\rm obs}$
represents the full-sky average of the effective number of
observations for each pixel.

\subsection{ILC7 and NILC5}
The WMAP instrument is composed of 10 DAs spanning five frequencies from 23 to 94 GHz \citep{bennett2003}.
The internal linear combination (ILC) method has been used by WMAP team to generate the WMAP ILC maps \citep{hinshaw2007,gold2011}.
The 7-year ILC (written as ``ILC7") map is a weighted combination from all five original frequency bands, which are smoothed to a common resolution of one degree.
For the 5-year WMAP data, in \citep{delabrouille2009} the authors have made a higher resolution CMB
ILC map (written as ``NILC5"), an implementation of a constrained linear combination of the channels with minimum error variance on a frame of spherical called needlets\footnote{The similar map for 7-year WMAP data is recently gotten in \citep{basak2011}.}.
In this paper, we will also consider both these ILC maps for the analysis. Note that all these WMAP data have the same resolution parameter $N_{\rm side}=512$, and the corresponding total pixel number $N_{\rm pix}=3 145 728$.

In comparison with WMAP observations to give constraints on the statistics, a ${\rm \Lambda CDM}$ cosmological model is assumed with the parameters given by the WMAP 7-year best-fit values \citep{komatsu2011}: $100\Omega_{\rm b}h^2=2.255$, $\Omega_ch^2=0.1126$, $\Omega_{\rm \Lambda}=0.725$, $n_s=0.968$, $\tau=0.088$ and $\Delta^2_{\mathcal{R}}(k_0)=2.430\times10^{-9}$ at $k_0=0.002{\rm Mpc}^{-1}$. We simulate the CMB maps for each frequency channel by considering the WMAP beam resolution and instrument noise, and then co-add them with inverse weight of the full-sky averaged pixel-noise variance in each frequency to get the simulated VW7 maps. Similar to the previous work \citep{hansen2012}, to simulate the ILC7 map, we ignore the noises and smooth the simulated map with one degree resolution. And for NILC5, we consider the noise level and beam window function given in \citep{delabrouille2009}. In all the random Gaussian simulations, we assume that the temperature fluctuations and instrument noise follow the Gaussian distribution, and do not consider any effect due to the residual foreground contaminations.

\section{Local properties of the CMB field}

\subsection{Local statistics}
In this section, we shall investigate the local properties of the CMB field, especially the WMAP Cold Spot, by using the local statistics: mean temperature, variance, skewness and kurtosis.

The statistics of local skewness and kurtosis were firstly
introduced in \citep{local1}. For a given CMB map with $N_{\rm
side}=512$ (VW7, ILC7 or NILC5), we degrade it to the lower
resolution $N_{\rm side}=256$ to reduce the effect of the noises.
And then, for this degraded map, the constructive process can be
formalized as follows: Let $\Omega(\theta_j,\phi_j;R)$ be a
spherical cap with an aperture of $R$ degree, centered at
$(\theta_j,\phi_j)$. We can define the functions $M$ (mean
temperature), $V$ (standard deviation), $S$ (skewness) and $K$
(kurtosis) that assign to the $j^{\rm th}$ cap, centered at
$(\theta_j,\phi_j)$ by the following way:
\begin{eqnarray}
M_j(R)&=&\frac{1}{N_{\rm p}}\sum_{i=1}^{N_{\rm p}} T_i, \nonumber\\
V_j(R)&=&\left(\frac{1}{N_{\rm p}}\sum_{i=1}^{N_{\rm p}} T^2_i\right)^{1/2},\nonumber \\
S_j(R)&=&\frac{1}{N_{\rm p}V_j^3}\sum_{i=1}^{N_{\rm p}} T^3_i,\nonumber\\
K_j(R)&=&\frac{1}{N_{\rm p}V_j^4}\sum_{i=1}^{N_{\rm p}} T^4_i-3,
\end{eqnarray}
where $N_{\rm p}$ is the number of pixels in the $j^{\rm th}$ cap,
$T_i$ is the temperature at $i^{\rm th}$ pixel. Obviously, the
values $S_j$ and $K_j$ obtained in this way for each cap can be
viewed as the measures of non-Gaussianity in the direction of the
center of the cap $(\theta_j,\phi_j)$. For a given aperture $R$,
we scan the celestial sphere with evenly distributed spherical
caps, and build the $M(R)$-, $V(R)$-, $S(R)$-, $K(R)$-maps. In our
analyses, we have chosen the locations of centroids of spots to be
the pixels in $N_{\rm side}=64$ resolution. By choosing different
$R$ values, one can study the local properties of the CMB field at
different scales. In \citep{local1, local2, local3}, the
statistics $S_j(R)$ and $K_j(R)$ with large $R$ values have been
applied to study the large-scale global non-Gaussianity in the CMB
field. However, in this paper we shall apply them to study the CMB
local properties.

\begin{figure}
\includegraphics[width=40mm]{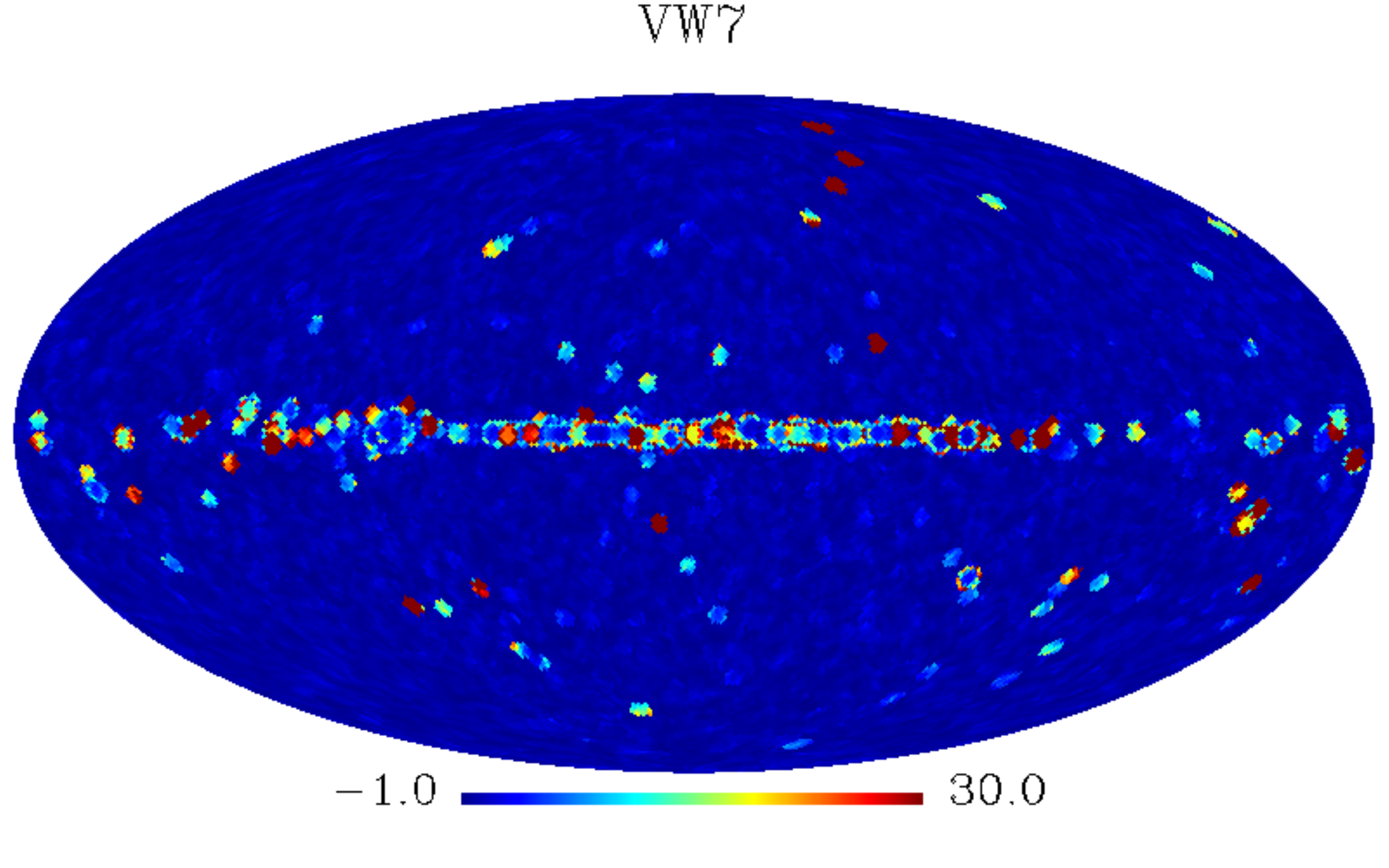}\includegraphics[width=40mm]{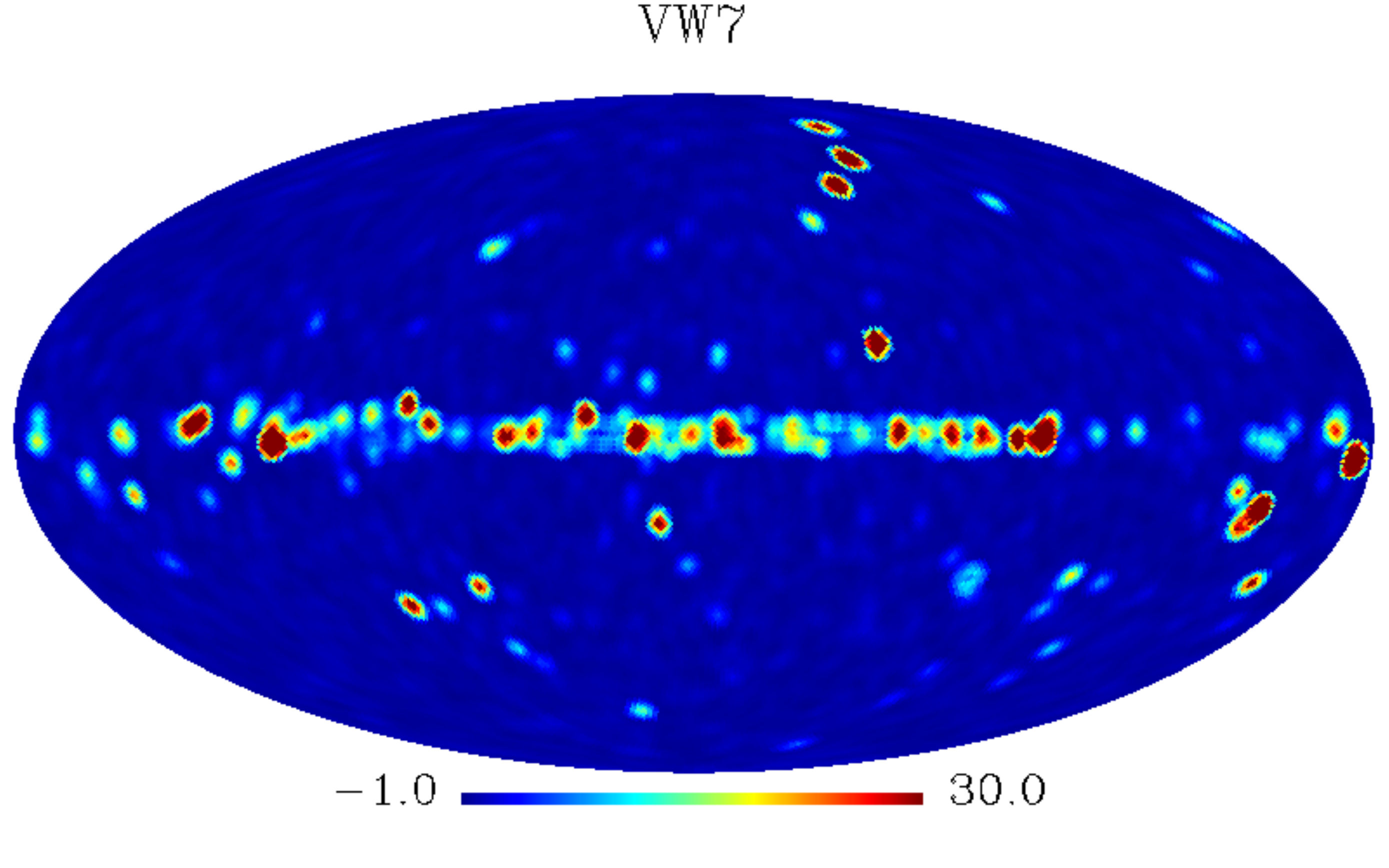}
    \caption{The $K(R)$ (left) and $\bar{K}(R)$ (right) maps for VW7 data. In both maps, we have adopted $R=2^{\circ}$.}
    \label{fig1}
\end{figure}

It is important to mention that these definitions cannot well
localize the non-Gaussian sources. For example, in Fig. \ref{fig1}
the kurtosis map $K(R)$ (left panel), we find the clear circular
morphology around the point sources. This means that the values of
$K_j$ always maximize/minimize at the edge of the circles, rather
than the center of the circles. To overcome it and localize the
non-Gaussian sources, it is better to define the following average
quantities,
\begin{eqnarray}
\bar{X_j}(R)=\frac{1}{N_{\rm p}}\sum_{j=1}^{N_{\rm p}} X_j(R),
\end{eqnarray}
where $X=M,V,S,K$ for mean temperature, standard deviation,
skewness and kurtosis. $X_j$ is the corresponding local quantities
defined above, and $N_{\rm p}$ is again the number of pixels in
the $j^{\rm th}$ cap. For the comparison, we plot the
corresponding $\bar{K}(R)$ in the right panel of Fig. \ref{fig1}.

\begin{figure}
\includegraphics[width=40mm]{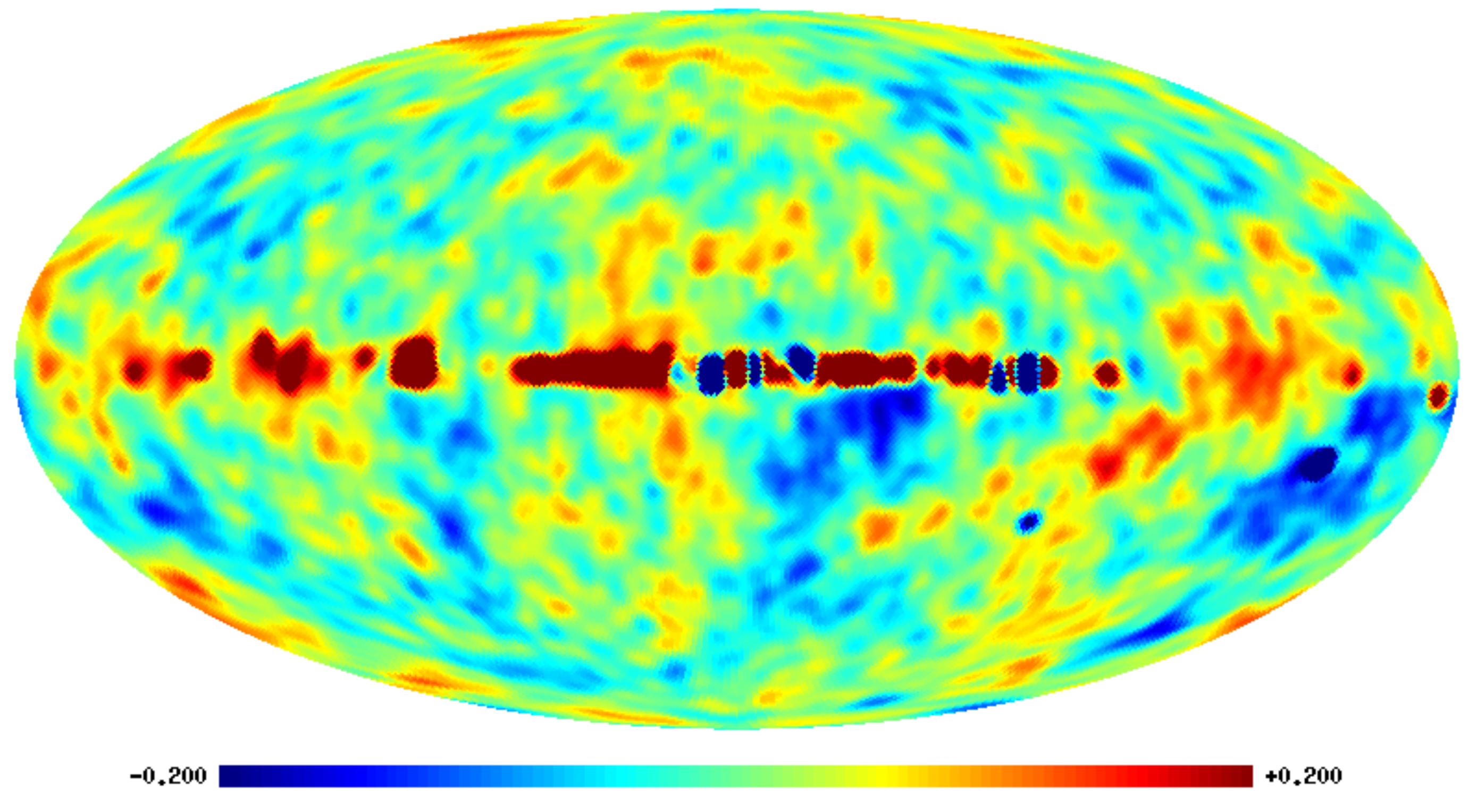}\includegraphics[width=40mm]{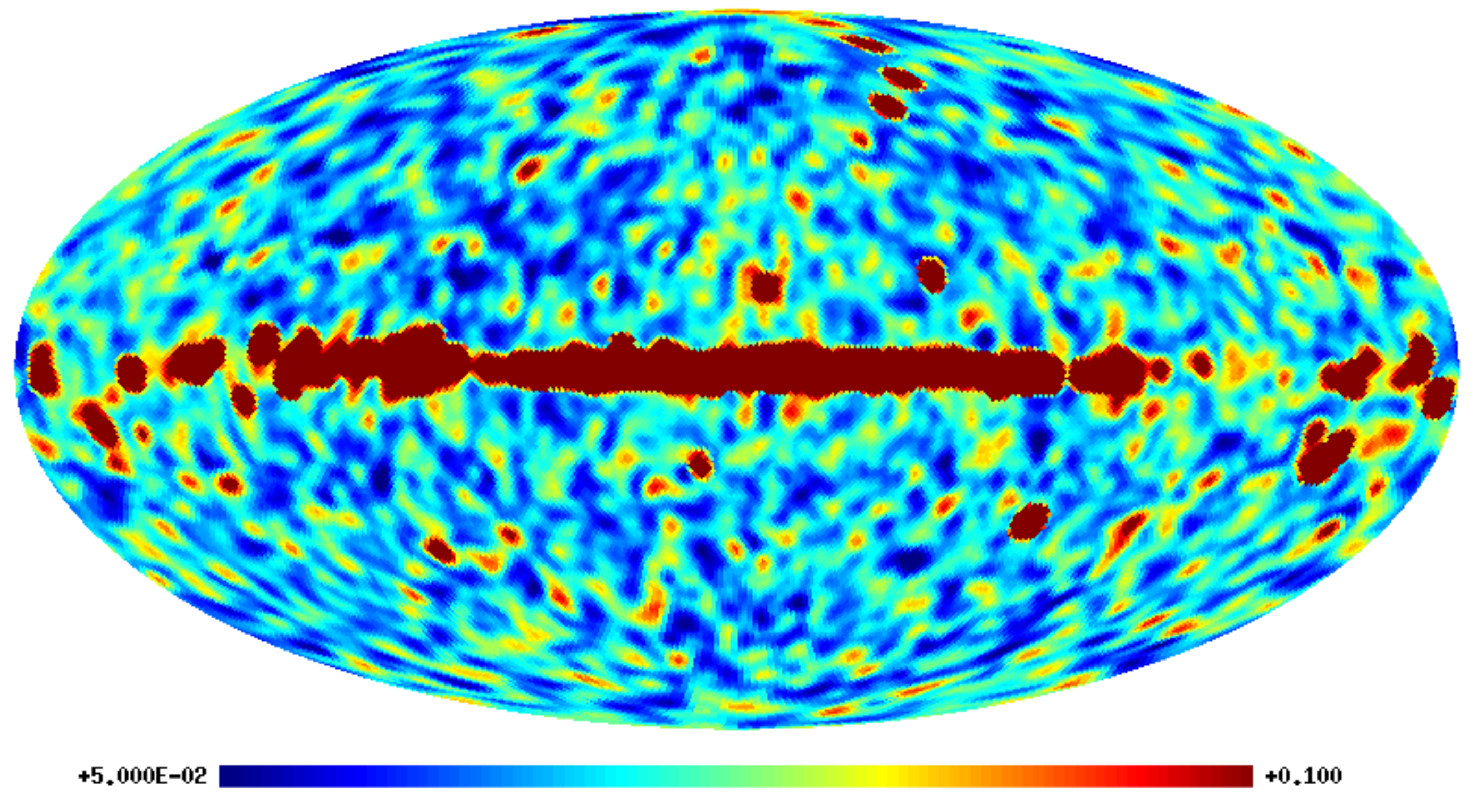}
\includegraphics[width=40mm]{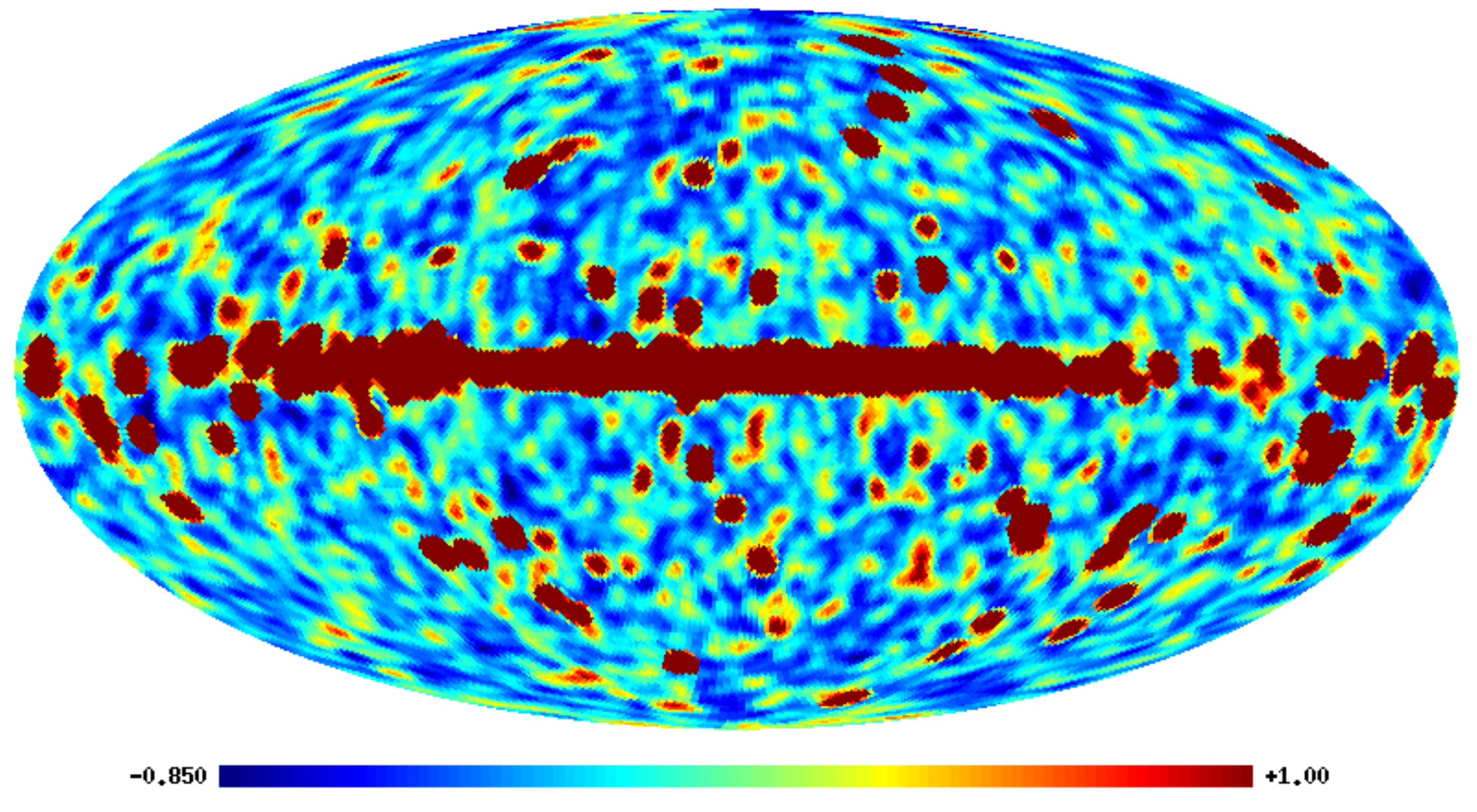}\includegraphics[width=40mm]{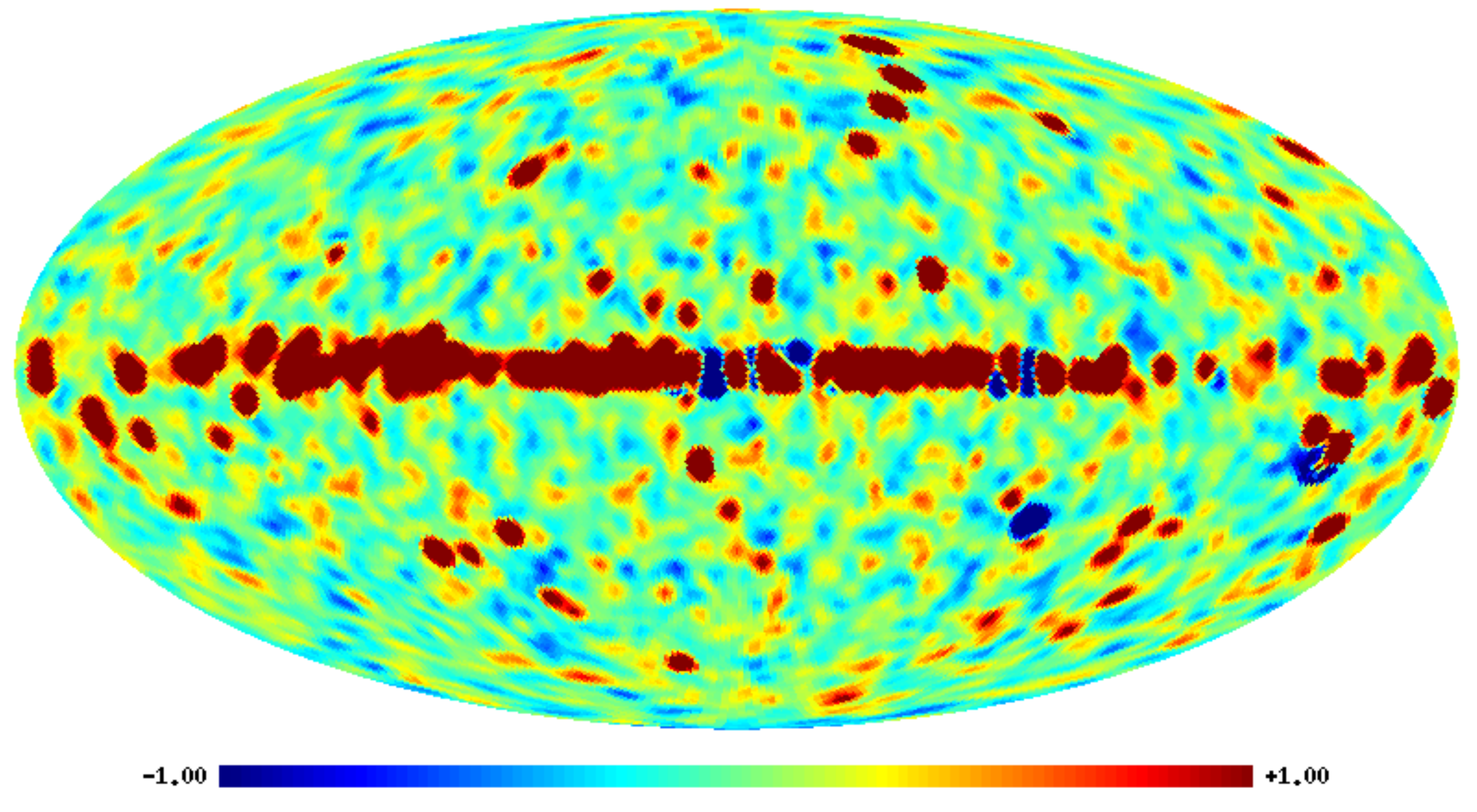}
    \caption{Clockwise, $\bar{M}(R)$ map, $\bar{V}(R)$ map, $\bar{S}(R)$ map and $\bar{K}(R)$ map for VW7,
where $R=2^{\circ}$. Note that the $\bar{M}(R)$ and $\bar{V}(R)$ maps have the unit: mK.}
    \label{fig2}
\end{figure}

\begin{figure}
\includegraphics[width=40mm]{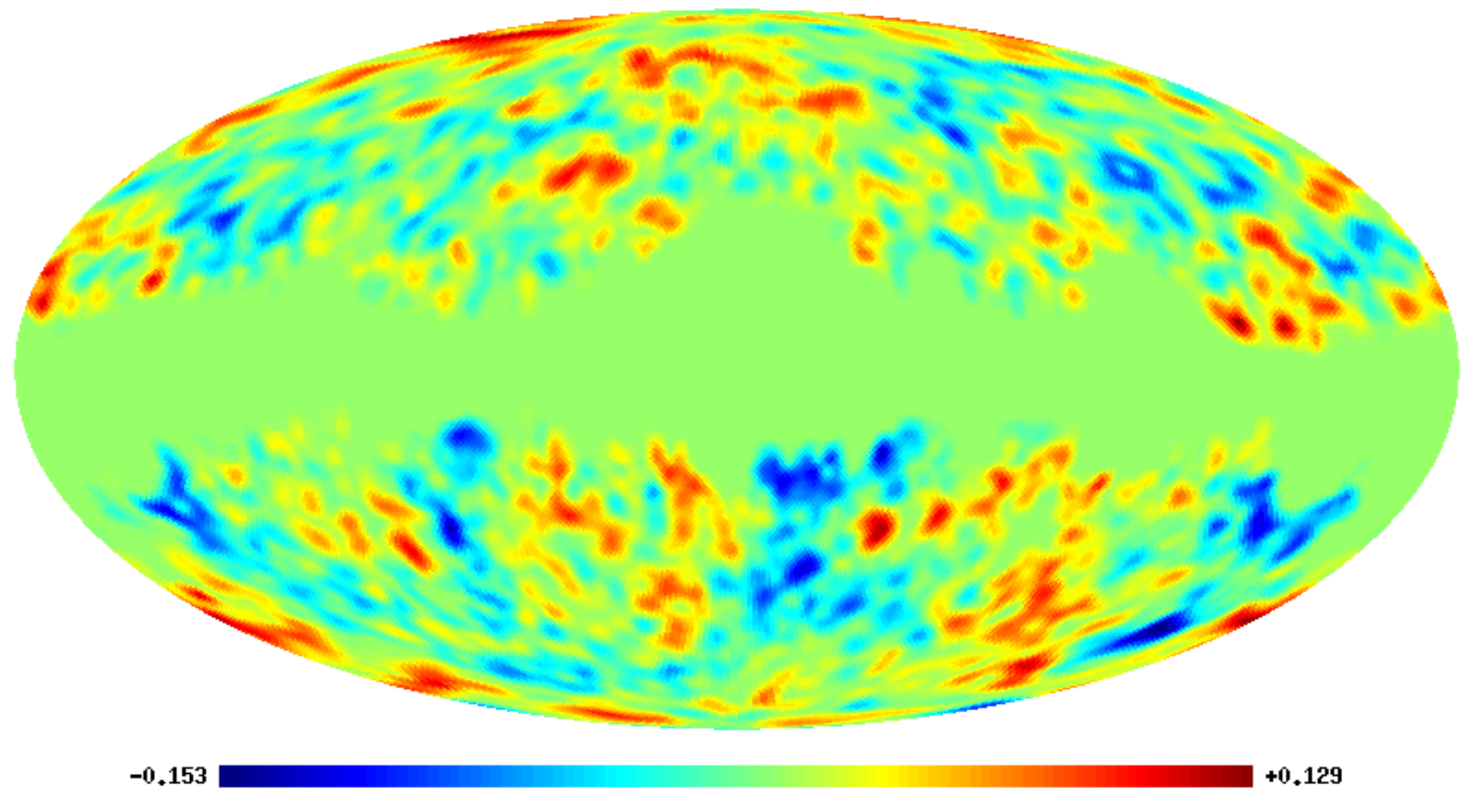}\includegraphics[width=40mm]{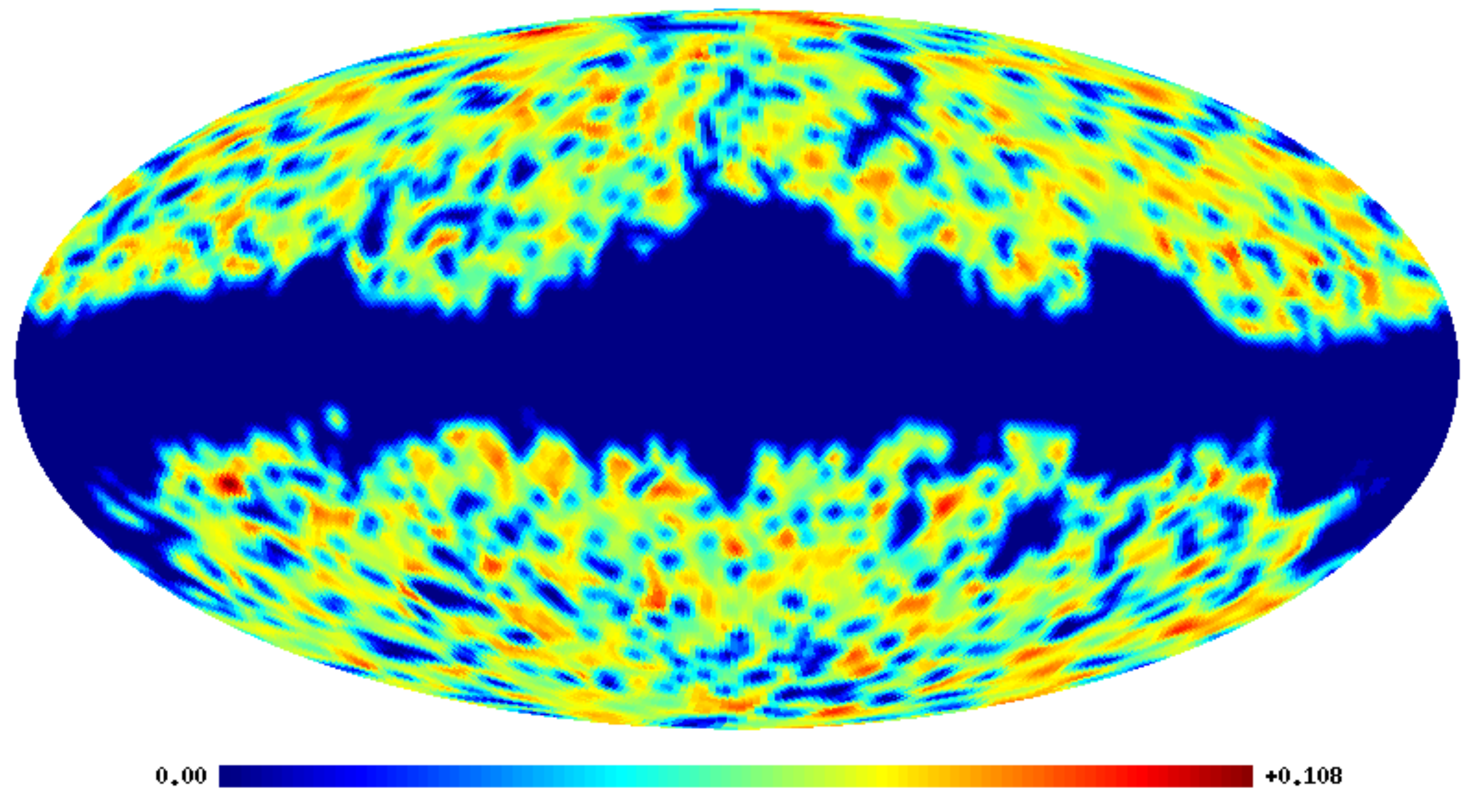}
\includegraphics[width=40mm]{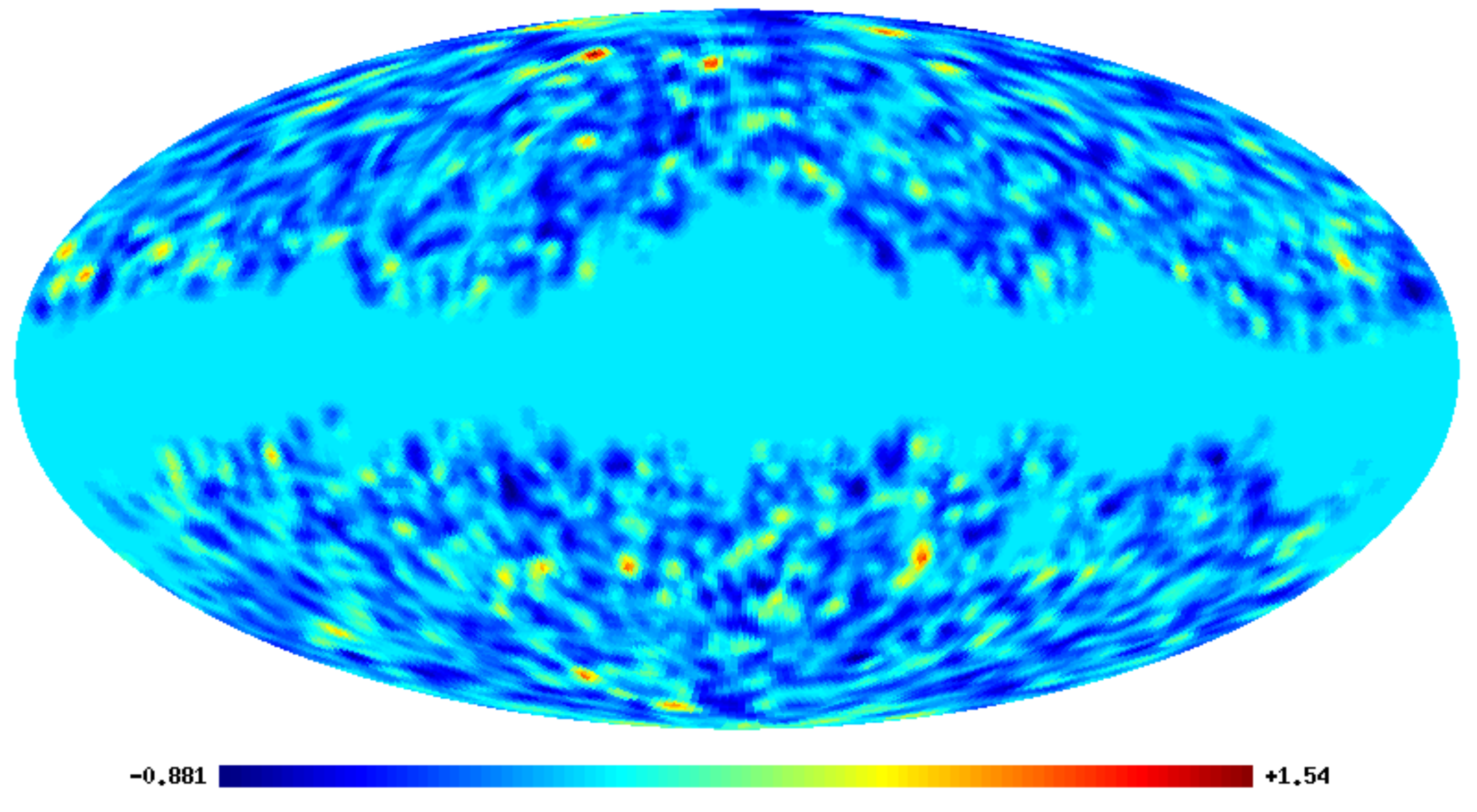}\includegraphics[width=40mm]{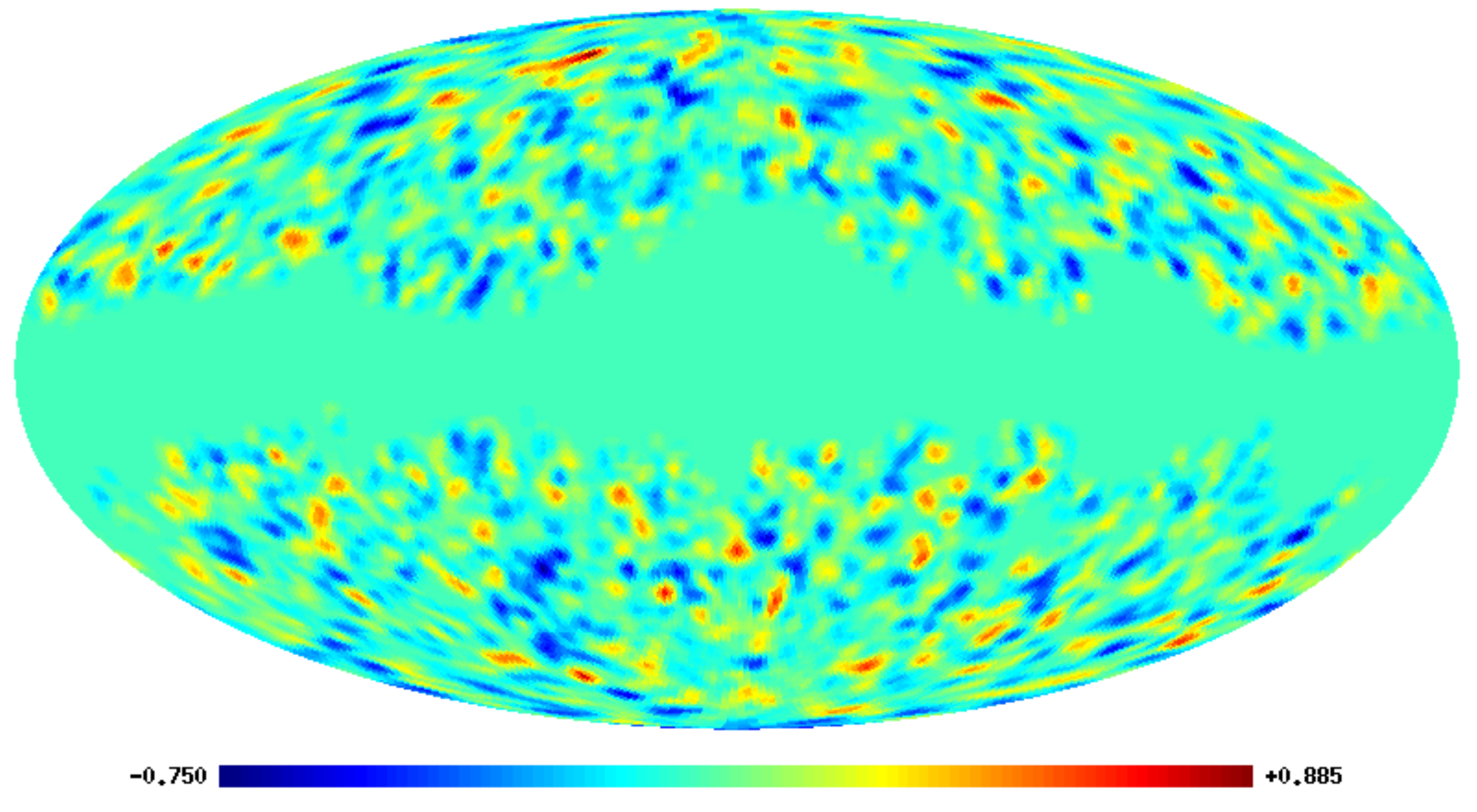}
    \caption{Same as Fig. \ref{fig2}, but KQ75y7 mask has been applied.}
    \label{fig22}
\end{figure}

Now, let us apply the method to the CMB maps. Firstly, we consider
the VW7 map. By choosing $R=2^{\circ}$, we plot $\bar{X}(R)$ maps
in Fig. \ref{fig2}. The figures clearly show that these local
statistics are sensitive to the foreground residuals and various
point sources. From $\bar{M}$-map, one finds that most
non-Gaussianities come from the Galactic plane around
$b=0^{\circ}$. However, from $\bar{V}$-, $\bar{S}$- and
$\bar{K}$-maps, various extra point sources far from the Galactic
plane are clearly presented. These contaminations can be well
excluded by the KQ75y7 mask \citep{gold2011}, which is clearly
shown in Fig. \ref{fig22}. In this figure, we plot the same
figures as those in Fig. \ref{fig2}, but the mask is applied.

\begin{figure}
\includegraphics[width=40mm]{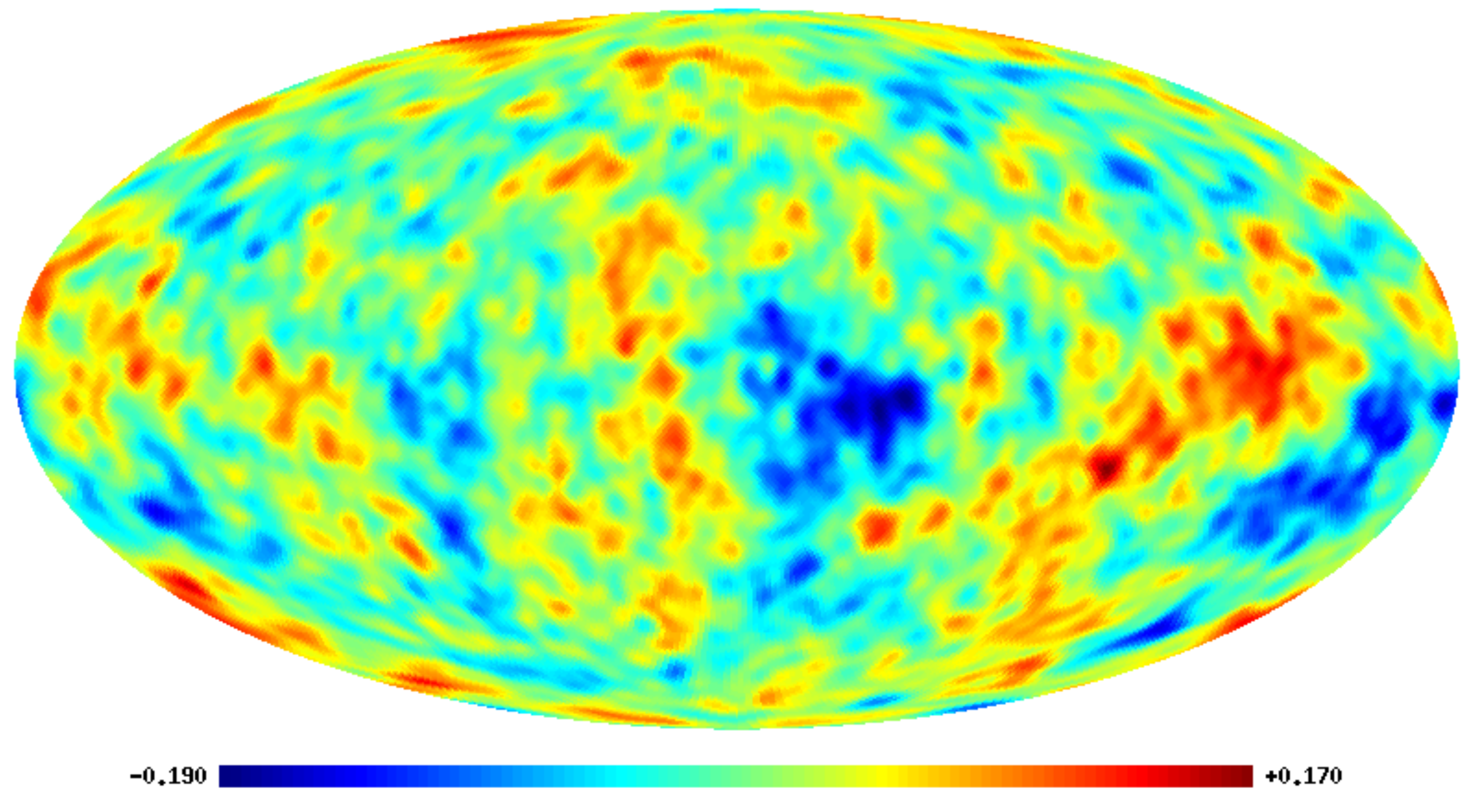}\includegraphics[width=40mm]{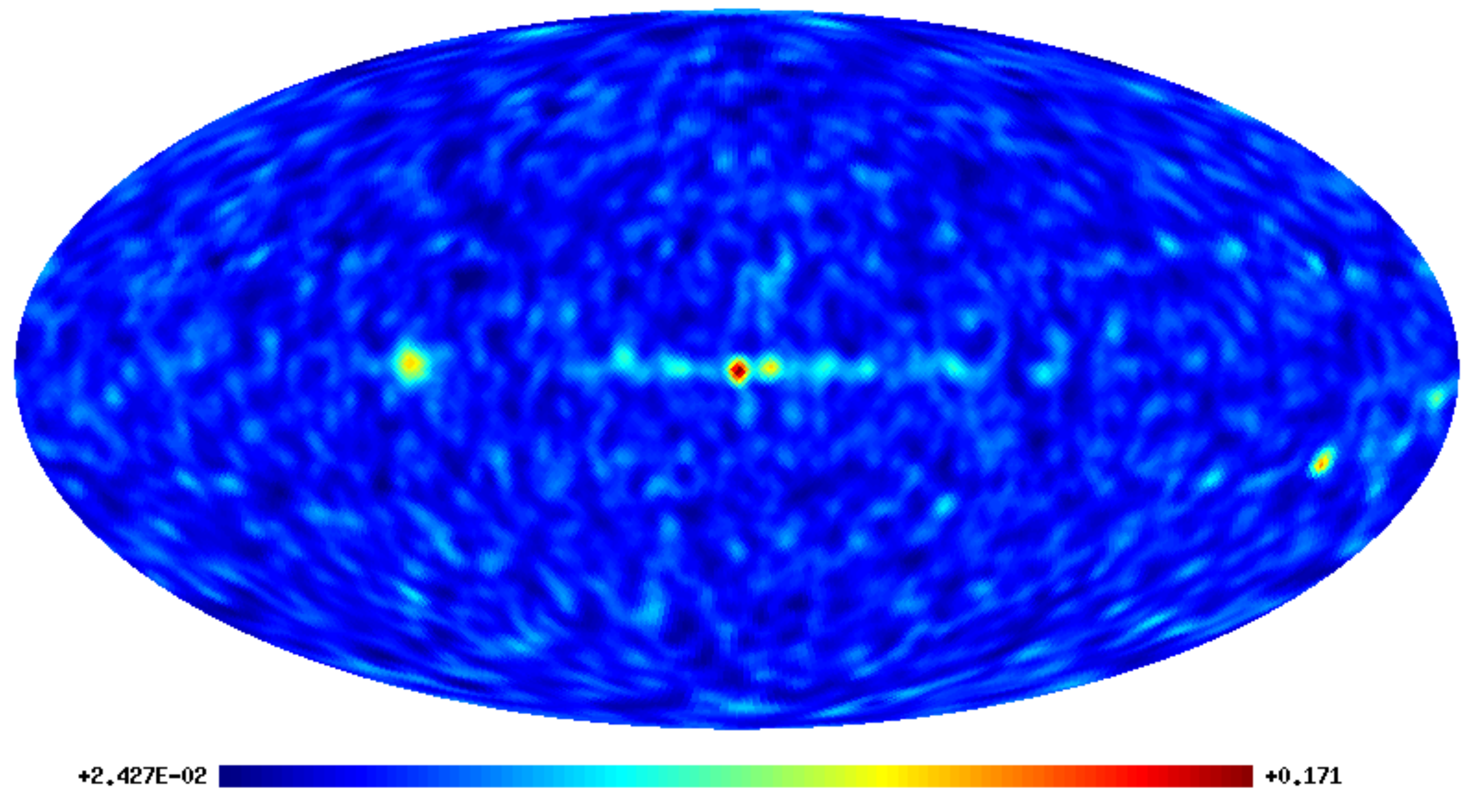}
\includegraphics[width=40mm]{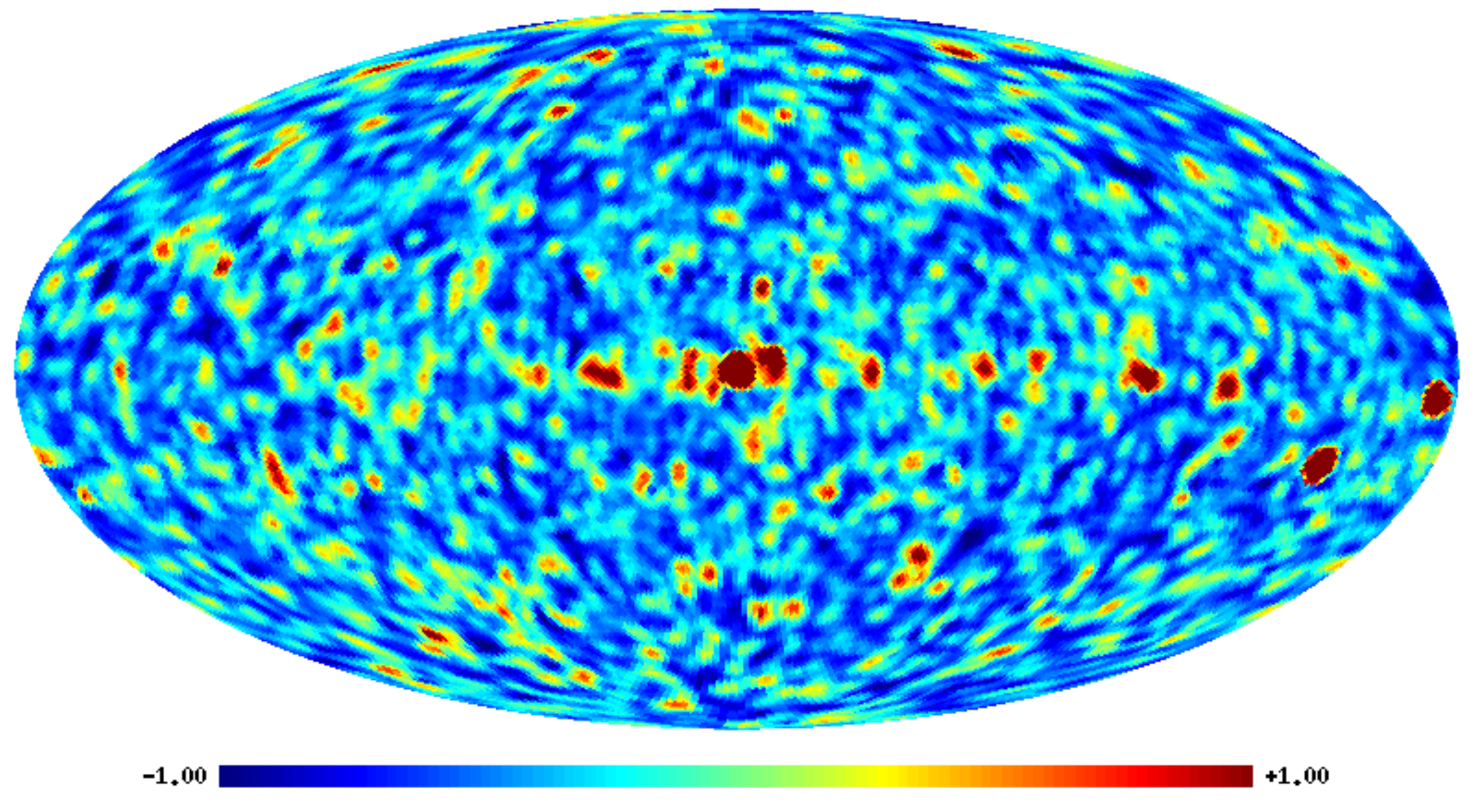}\includegraphics[width=40mm]{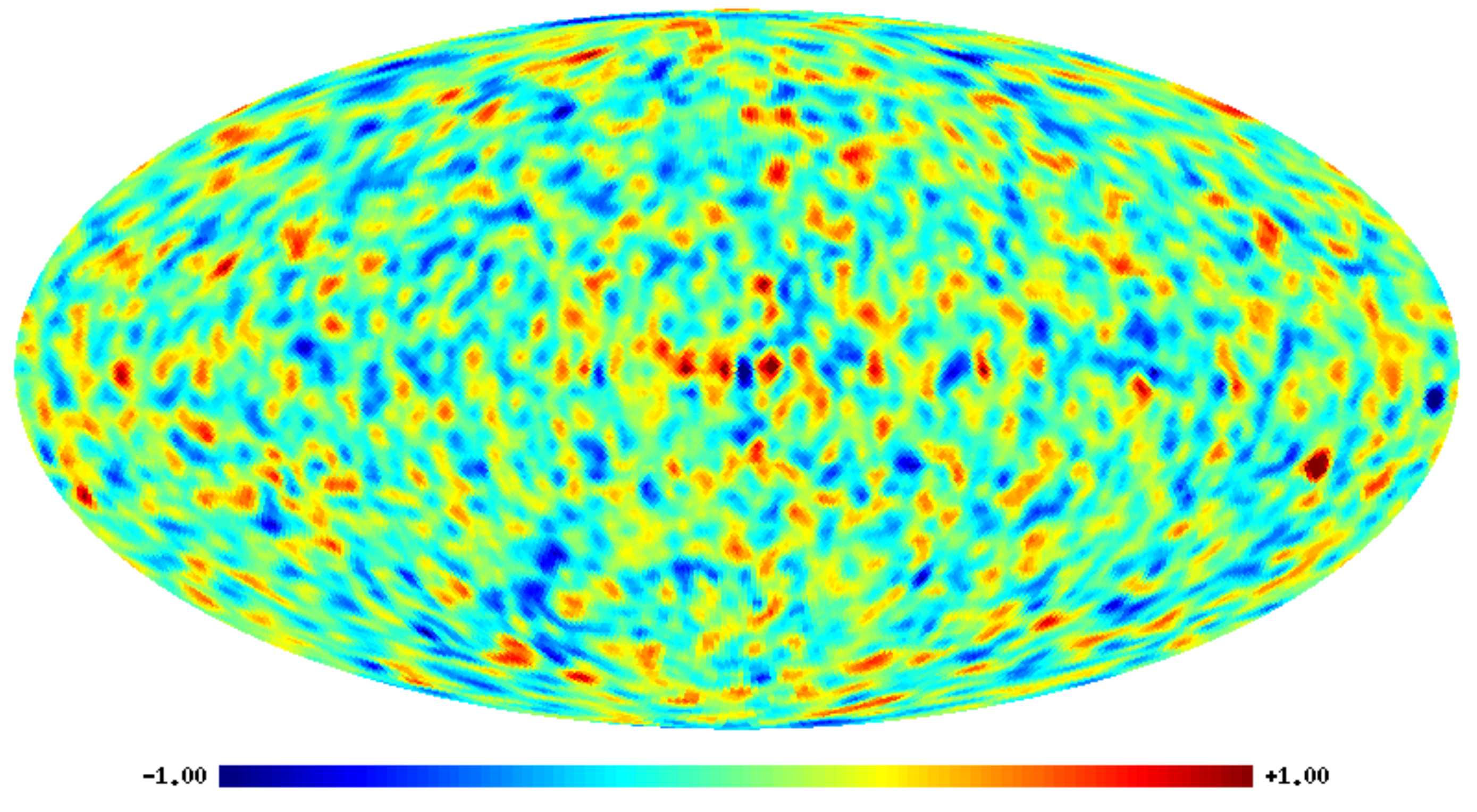}
    \caption{Same as Fig. \ref{fig2}, but VW7 map is replaced by ILC7 map.}
    \label{fig3}
\end{figure}

\begin{figure}
\includegraphics[width=40mm]{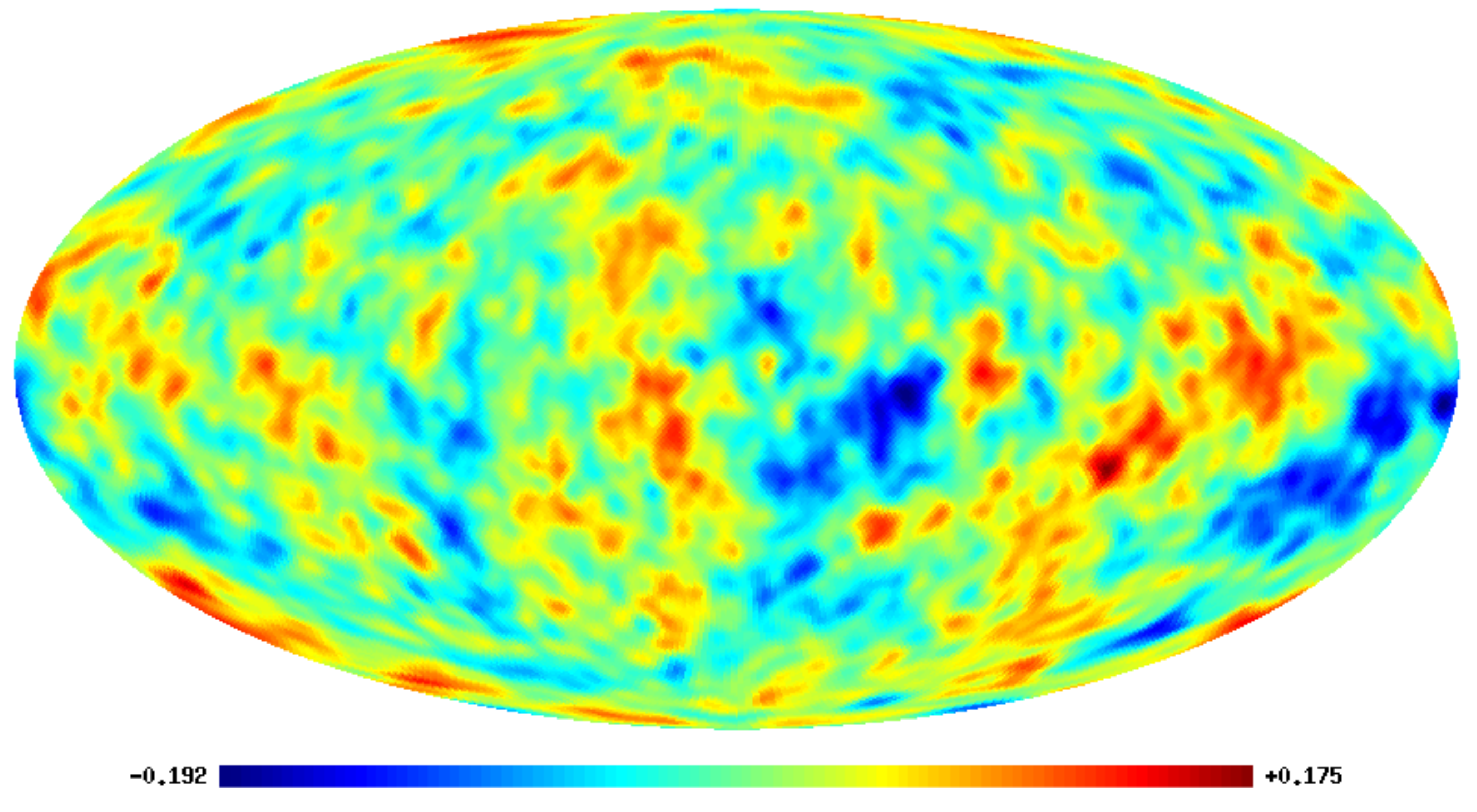}\includegraphics[width=40mm]{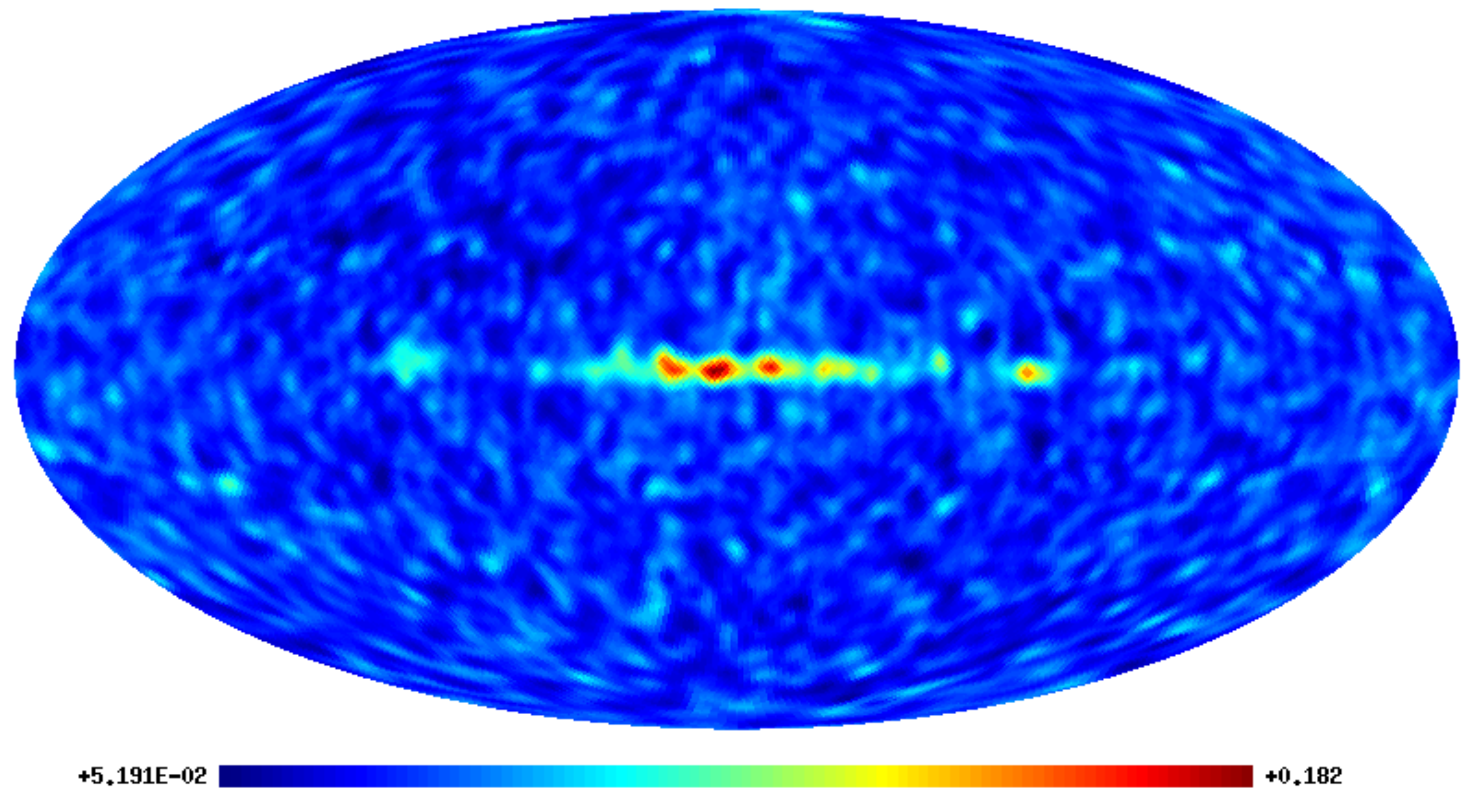}
\includegraphics[width=40mm]{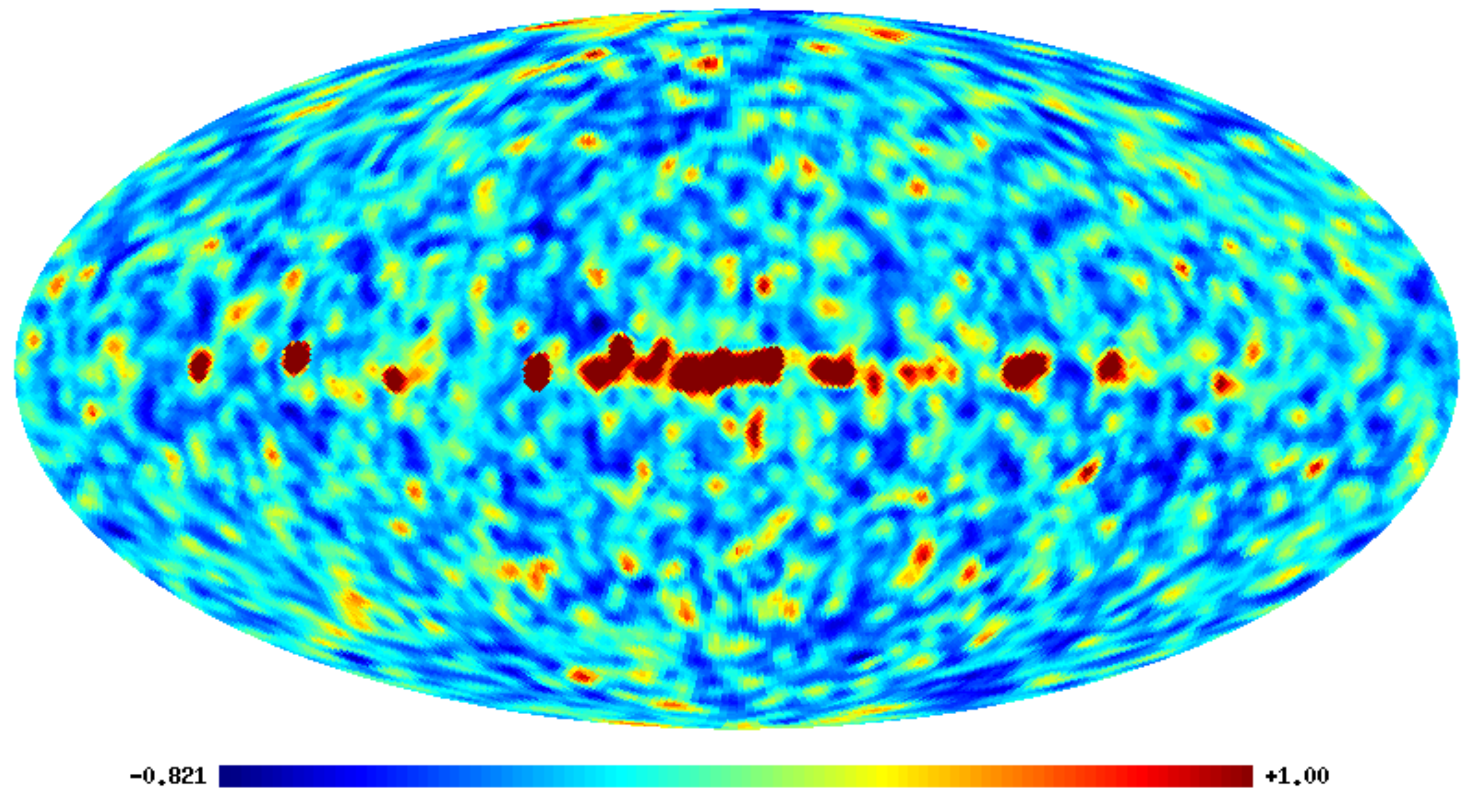}\includegraphics[width=40mm]{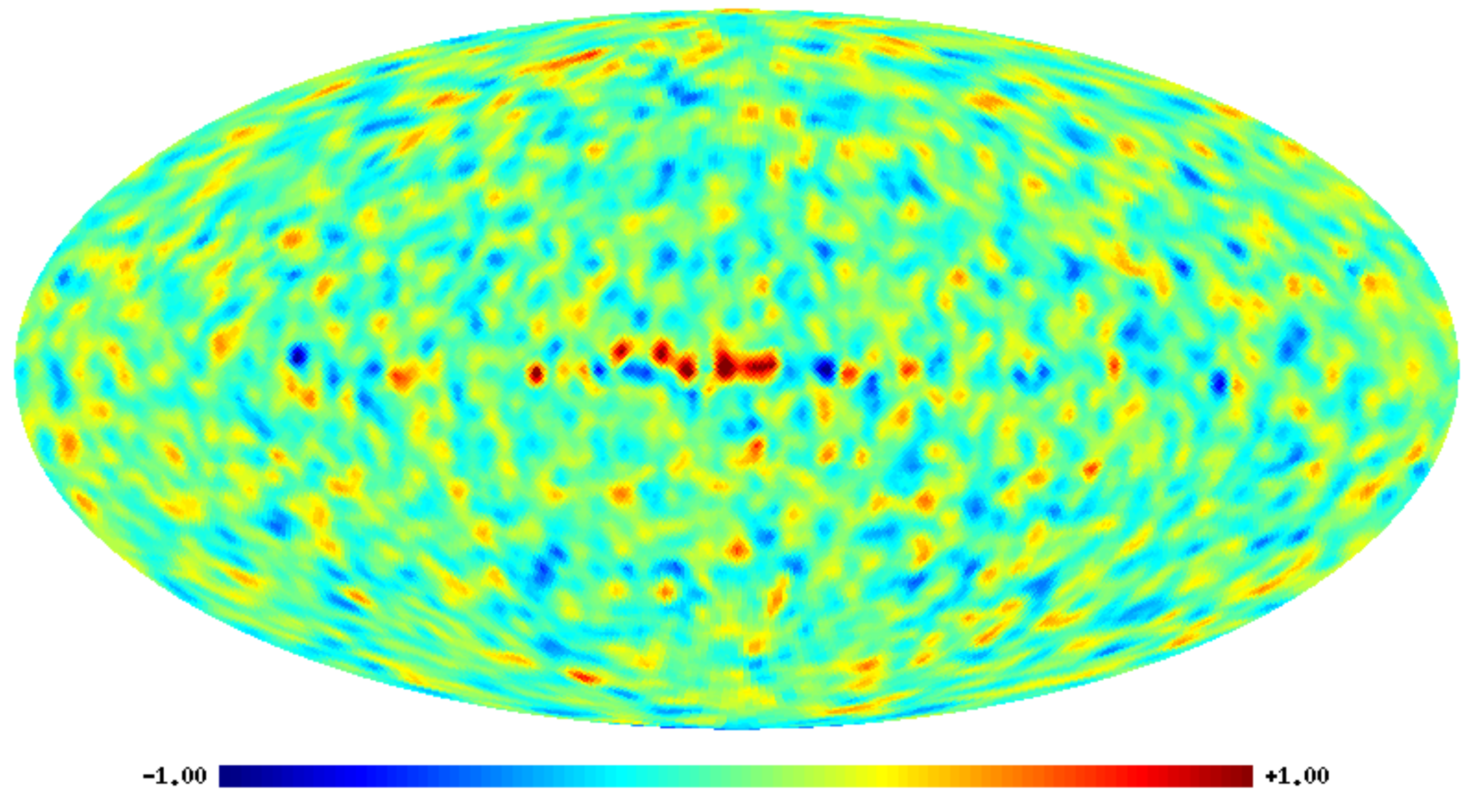}
    \caption{Same as Fig. \ref{fig2}, but VW7 map is replaced by NILC5 map.}
    \label{fig4}
\end{figure}

Similarly, we also apply the method to ILC7 and NILC5 maps by
choosing $R=2^{\circ}$. The results are shown in Fig. \ref{fig3}
and Fig. \ref{fig4}. We find that these ILC maps are much cleaner
than VW7 map in all the four $\bar{M}$-, $\bar{V}$-, $\bar{S}$-,
$\bar{K}$-maps. Even so, from the $\bar{V}$-, $\bar{S}$-,
$\bar{K}$-maps, we also find some non-Gaussian sources in the
Galactic plane. In addition, two significant sources at
($l=209.5^{\circ},b=-20.1^{\circ}$) and
($l=184.9^{\circ},b=-5.98^{\circ}$) are clearly presented in ILC7
maps, which have been identified as the known point sources, and
excluded by the KQ75y7 mask. NILC5 map is slightly clearer than
ILC7, especially the significant point sources at
($l=209.5^{\circ},b=-20.1^{\circ}$) and
($l=184.9^{\circ},b=-5.98^{\circ}$) disappear now. But the
contaminations in the Galactic plane are still quite significant.

\subsection{The local properties of WMAP Cold Spot}
\begin{figure}
\includegraphics[width=80mm]{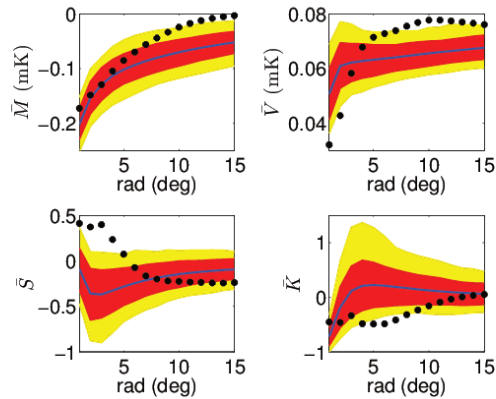}
    \caption{Four statistics for the \emph{coldest spots}. Confidence regions obtained from 500 Monte Carlo simulations are shown for 68 per cent (dark central region, red online) and
95 per cent (light outer region, yellow online) levels, as is the mean (solid blue line). The observed statistics for WMAP ILC7 map are shown by the solid dots (black online).}
    \label{fig5}
\end{figure}

In this subsection, we shall focus on the local statistics for
WMAP CS, and compare with those of the \emph{coldest spot} in
random Gaussian simulations. For a given $\bar{X}(R)$ map
($R=M,V,S$ or $K$) derived from WMAP data, the values of
$\bar{X}(R)$ centered at CS are calculated for the scales of
$R=1^{\circ}$, $2^{\circ}$, $3^{\circ}$, $4^{\circ}$, $5^{\circ}$,
$6^{\circ}$, $7^{\circ}$, $8^{\circ}$, $9^{\circ}$, $10^{\circ}$,
$11^{\circ}$, $12^{\circ}$, $13^{\circ}$, $14^{\circ}$,
$15^{\circ}$. From Fig. \ref{fig2}, we find in the maps derived
from VW7 data, there are many point sources. So, for a fair
comparison, in this subsection we shall only consider the ILC7 and
NILC5 maps. The statistics for the ILC7 maps are displayed in Fig.
\ref{fig5}. We compare them with 500 Gaussian simulations.

For each simulated temperature anisotropy map $\Delta T(\hat{n})$ with $N_{\rm side}=512$, we search for the
\emph{coldest spot} and its position $(l_c, b_c)$, which will be used for the comparison. By the
exactly same process, we derive the corresponding
$\bar{X}(R)$ maps. Then, for each $X$ and $R$, we study the
distribution of 500 $\bar{X}_{c}(R)$ values ($\bar{X}_{c}(R)$ is the
statistic of the \emph{coldest spot} in the corresponding
simulation), and construct the confident intervals for the
statistic. The $68\%$ and $95\%$ confident intervals are
illustrated in Fig. \ref{fig5}.

As we can imagine, if CS is simply cold without any other
non-Gaussianity, the statistics for $\bar{V}$,
$\bar{S}$ and $\bar{K}$ should be normal, i.e. close to the mean
values of Gaussian simulations for any $R$. On the other hand, if
CS is a combination of some small-scale non-Gaussian structures,
as some explanations in \citep{cruz2006}, the local variance,
skewness and kurtosis in small scales should be quite large.
However, as we will show below, none of these is the case of WMAP
CS.

From Fig. \ref{fig5}, we find that for $\bar{M}$ statistic, WMAP
CS is excellently consistent with Gaussian simulations when $R\le
5^{\circ}$. However, when $R>10^{\circ}$, it deviates from
simulations at more than $95\%$ confident level. This is caused by
the fact that WMAP CS is surrounded by an anomalous hot ring-like
structure, which is firstly noticed by Zhang \& Hunterer in
\citep{zhang2010}. For $\bar{V}$ statistic, deviations from
Gaussianity outside the $95\%$ confident regions are at the scales
$R<2^{\circ}$ and $R>6^{\circ}$. Furthermore, the deviations
outside the $95\%$ confident regions are detected in skewness at
scales of $R<5^{\circ}$ and in kurtosis at scales of
$4^{\circ}<R<7^{\circ}$. For the NILC5 map, the similar deviations
for these statistics have also been derived.

Combining these results, we find that WMAP CS seems to be a nontrivial large-scale structure, rather than a
combination of some small non-Gaussian structures (for instance,
the point sources or foreground residuals, which always follow
the non-Gaussianity in the small scales as shown in Fig.
\ref{fig2}). This is one of the main conclusions of this paper.

\begin{figure}
\includegraphics[width=80mm]{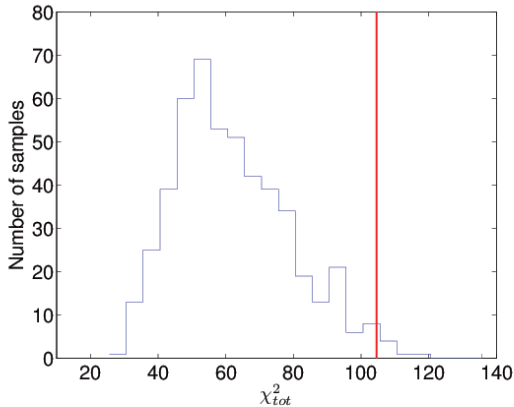}
    \caption{Histogram of the $\chi^2_{tot}$ statistic for the \emph{coldest spots} obtained from 500 Monte Carlo simulations. The observed statistic for WMAP ILC7 map is shown by the solid vertical line (red online).}
    \label{fig6}
\end{figure}

We now consider, in more details, the most significant deviation
from Gaussianity obtained in Fig. \ref{fig5}. Similar to
\citep{mcewen2005}, for each panel of Fig. \ref{fig5}, we define
the $\chi^2$ statistic as follows:
  \begin{equation}
  \chi^{2}_{Y}=\sum_{R_i,R_j} (Y_0(R_i)-\overline{Y}(R_i))\Sigma^{-1}_{ij}(Y_0(R_j)-\overline{Y}(R_j)), \label{chi2}
 \end{equation}
where $Y=\bar{M},\bar{V},\bar{S},\bar{K}$. $R_i$ and $R_j$ run
through $1^{\circ}$ to $15^{\circ}$. $Y_0(R_i)$ are the values of
the statistics for WMAP CS, and $Y(R_i)$ are those for the
simulations. $\overline{Y}(R_i)$ is average value of $Y(R_i)$.
$\Sigma$ is the covariance matrix of the vector $Y(R_i)$. Note
that the correlations between $\bar{M}(R_i)$ and $\bar{M}(R_j)$
($i\neq j$) are very strong (the corresponding correlation
coefficienta are all larger than 0.6), which significantly affect the
corresponding $\chi^{2}_{\bar{M}}$ value, especially when the values of $Y_0(R_i)-\overline{Y}(R_i)$ oscillate for different $R_i$.
The total value can also be defined as
$\chi^2_{tot}=\chi^{2}_{\bar{M}}+\chi^{2}_{\bar{V}}+\chi^{2}_{\bar{S}}+\chi^{2}_{\bar{K}}$.
We list the $\chi^{2}_{Y}$ values (Case 1) in Table \ref{table1}
for ILC7 and in Table \ref{table2} for NILC5. In order to be compared with
Gaussian simulations, for each realization, we repeat the
calculation in Eq.(\ref{chi2}), but the quantities $Y_0(R_i)$ of
WMAP CS are replaced by the corresponding quantities of the
Gaussian realization. Fig. \ref{fig6} illustrates the histogram of
$\chi^2_{tot}$ statistic for the ILC7 based map, where we find
that WMAP CS in ILC7 deviates from Gaussianity at the $99.0\%$
significant level. At the same time, we also obtain the same results from NILC5 map.

\begin{table}

 \centering

 \begin{minipage}{80mm}

  \caption{The $\chi^2$ values of various statistics for ILC7 based maps.
In Case 1, WMAP CS compares with the \emph{coldest spots} in 500 random Gaussian simulations,
in Case 2, WMAP CS compares with the spots at ($l=209^{\circ}$,$b=-57^{\circ}$)  in 500 simulations,
and Case 3 is same with Case 2, but a cosmic texture has been superimposed in each simulated sample. \label{table1}}

  \begin{tabular}{@{}llrrrrlrlr@{}}

  \hline
     & $\chi^2_{\bar{M}}$ & $\chi^2_{\bar{V}}$ & $\chi^2_{\bar{S}}$ & $\chi^2_{\bar{K}}$ & $\chi^2_{tot}$\\
\hline
Case 1     &  32.95    &  34.95   &  24.13   & 12.64    & 104.66    \\
Case 2     &  35.16    &  43.27   &  31.83   & 10.46    & 120.73    \\
Case 3     &  34.01    &  29.04   &  12.76   & 7.40     & 83.21     \\
\hline

\end{tabular}

\end{minipage}

\end{table}

\begin{table}

 \centering

 \begin{minipage}{80mm}

  \caption{Same as Table \ref{table1}, but for NILC5 maps.\label{table2}}

  \begin{tabular}{@{}llrrrrlrlr@{}}

  \hline
     & $\chi^2_{\bar{M}}$ & $\chi^2_{\bar{V}}$ & $\chi^2_{\bar{S}}$ & $\chi^2_{\bar{K}}$ & $\chi^2_{tot}$\\
\hline
Case 1  &  31.59    &  26.27   &  35.42   & 11.73    & 105.02    \\
Case 2  &  33.67    &  36.23   &  43.24   & 13.54    & 126.68    \\
Case 3  &  31.17    &  25.27   &  19.82   & 11.03    & 87.29     \\
\hline

\end{tabular}

\end{minipage}

\end{table}

\subsection{Compared with simulations including cosmic texture}
In \citep{cruz2007b,cruz2008}, by studying the temperature and
area of CS, the authors found that the cosmic texture, rather than
the other explanations, provided an excellent interpretation for
the WMAP CS\footnote{In the review paper
\citep{vielva2010}, one can find some other explanations, and the
corresponding criticisms, which will not be considered in this
paper.}. In this subsection, we shall study whether the local anomalies
of CS found in this work are consistent with the cosmic texture
explanation.

\begin{figure}
\includegraphics[width=80mm]{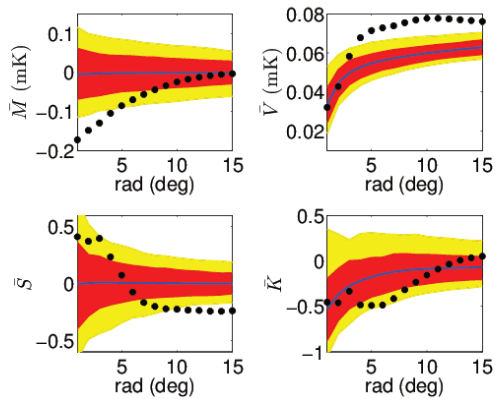}
    \caption{Four statistics for the spot at $(l=209^{\circ}, b=-57^{\circ})$. Confidence regions obtained from 500 Monte Carlo simulations are shown for 68 per cent (dark central region, red online) and
95 per cent (light outer region, yellow online) levels, as is the mean (solid blue line). The observed statistics for WMAP ILC7 map are shown by the solid dots (black online).}
    \label{fig55}
\end{figure}

\begin{figure}
\includegraphics[width=80mm]{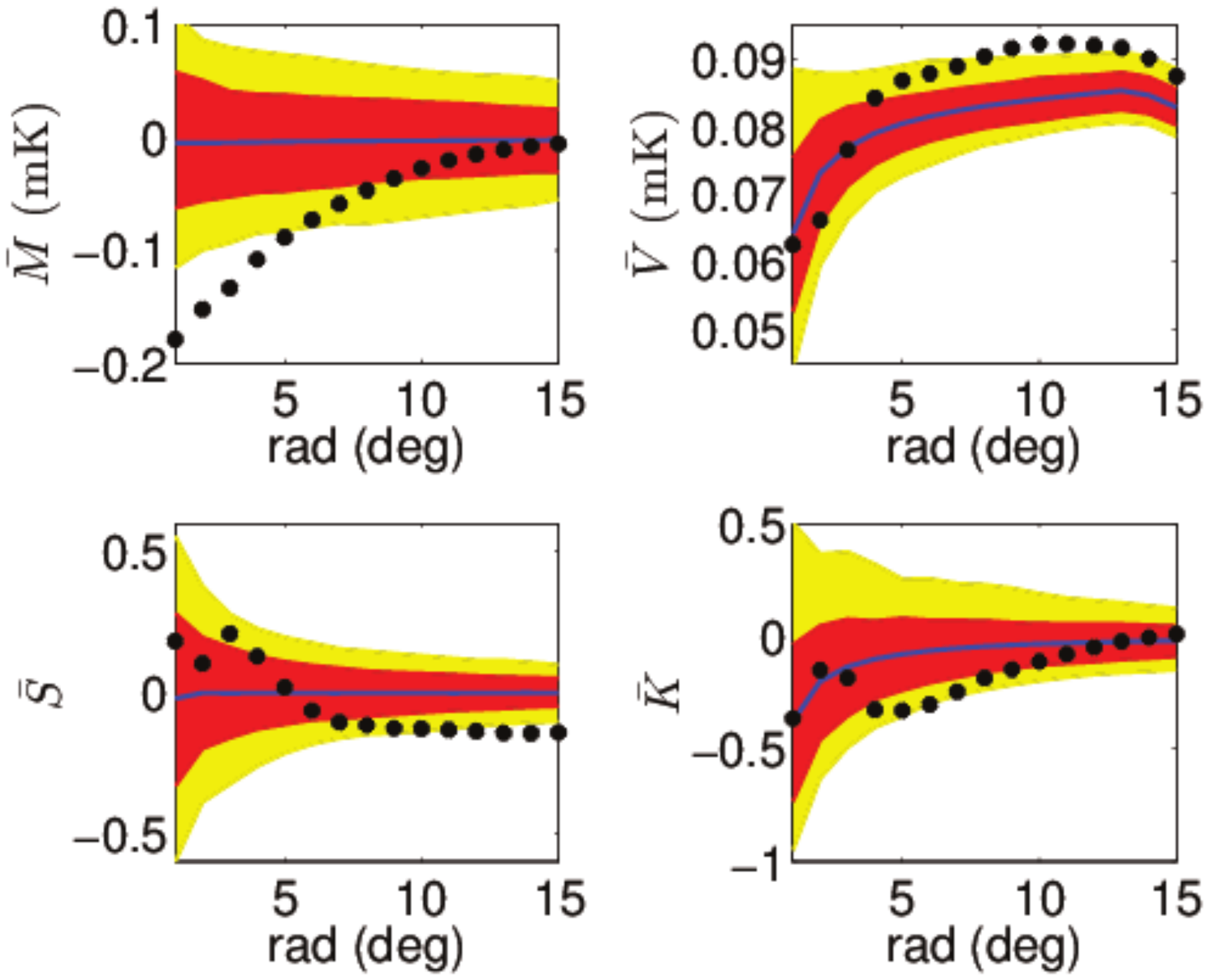}
    \caption{Same as Fig. \ref{fig55}, but ILC7 map is replaced by the masked VW7 maps.}
    \label{fig555}
\end{figure}

We firstly study how the WMAP CS deviates from the \emph{normal
spot} of the CMB map. We compare the CS with the spots at
$(l=209^{\circ}, b=-57^{\circ})$ in the Gaussian random
simulations. The statistics are displayed in Fig. \ref{fig55},
with the confidence intervals constructed from the Monte Carlo
simulations. As anticipated, we find that the CS is colder than
simulations in the small scales. Especially, when $R\le 4^{\circ}$,
it deviates from simulations at more than $95\%$ confident level.
However, as $R$ increases, the deviation becomes smaller and
smaller. Furthermore, we find that for the $\bar{V}$ and $\bar{S}$
statistics, WMAP CS deviates from simulations at larger scales,
i.e. it deviates at more than $95\%$ confident level when
$R>4^{\circ}$ for the $\bar{V}$ statistic and when $R>12^{\circ}$
for the $\bar{S}$ statistic. The similar results are also obtained
from NILC5 map and the masked VW7 map (see Fig. \ref{fig555}).

\begin{figure}
\includegraphics[width=80mm]{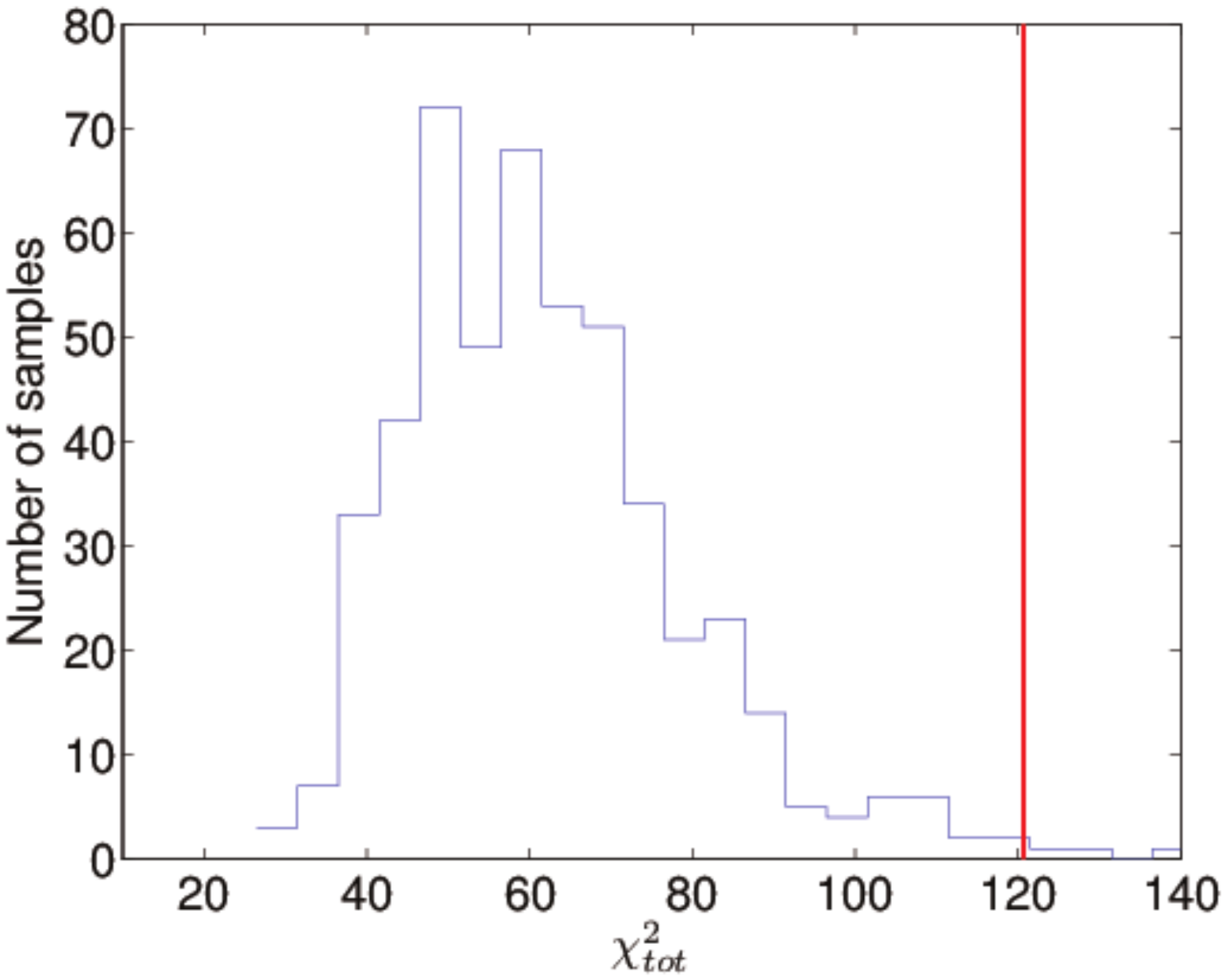}
    \caption{Histograms of the $\chi^2_{tot}$ statistic for the spots at $(l=209^{\circ}, b=-57^{\circ})$ obtained from 500 Monte Carlo simulations. The observed statistic for WMAP ILC7 map is shown by the solid vertical line (red online).}
    \label{fig66}
\end{figure}

\begin{figure}
\includegraphics[width=80mm]{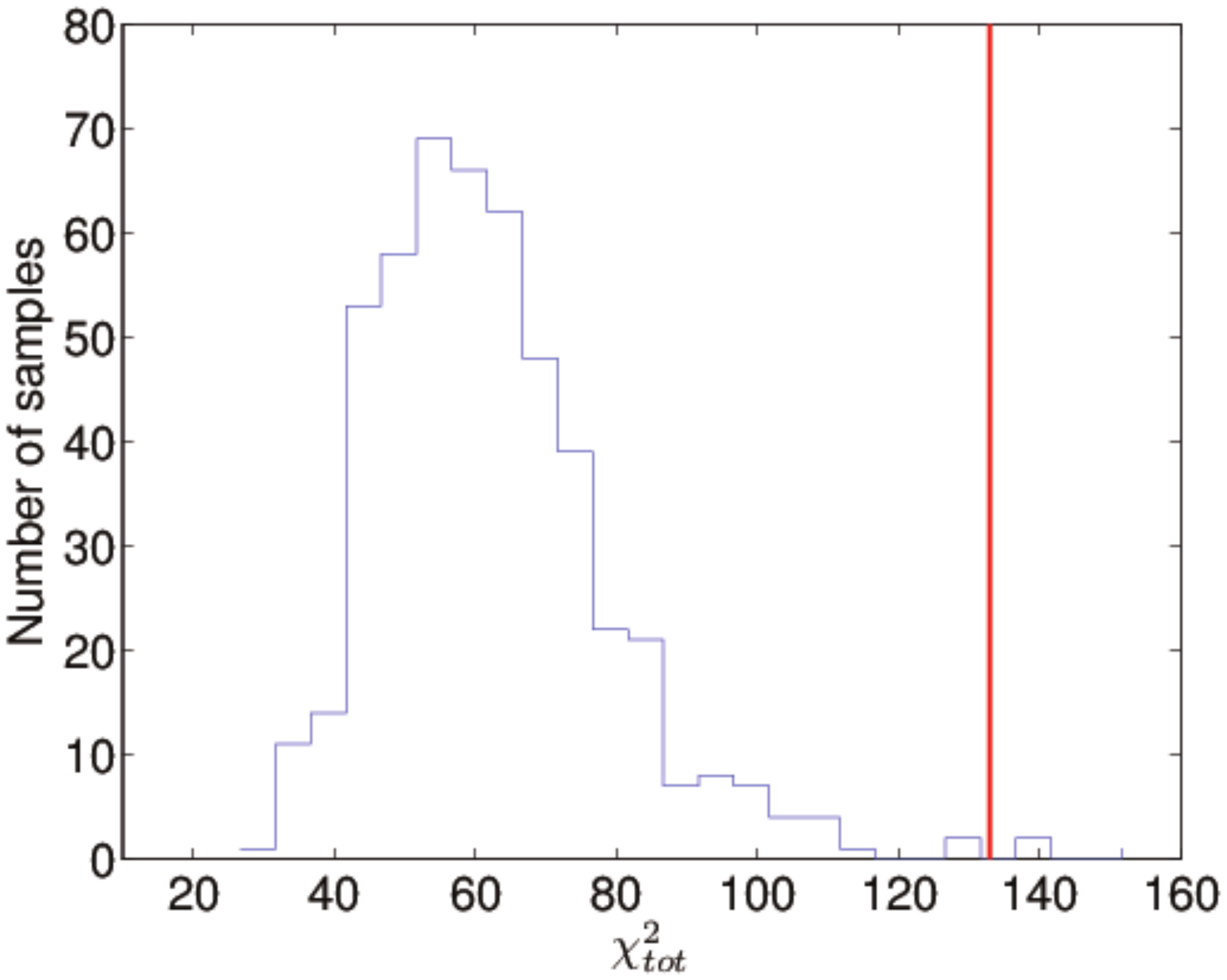}
    \caption{Same as Fig. \ref{fig66}, but ILC7 map is replaced by the masked VW7 maps.}
    \label{fig666}
\end{figure}

The $\chi^2$ statistic defined in Eq.(\ref{chi2}) is also applied
here. In Tables \ref{table1}, \ref{table2} and \ref{table3} (Case
2), we list the values of $\chi_{Y}^2$ and $\chi_{tot}^2$ for
ILC7, NILC5 and masked VW7, respectively. Compared with the
Gaussian simulations, ILC7 CS deviates from Gaussianity at
the $99.0\%$ significant level (see Fig. \ref{fig66}), NILC5 CS
deviates at the $89.8\%$ significant level, and the masked VW7 CS
deviates at the $99.4\%$ significant level (see Fig.
\ref{fig666}).

Now, let us study the cosmic texture interpretation. The profile for the CMB temperature fluctuation caused by a collapsing cosmic texture is given by
\begin{equation}
\label{eq:text_profile2}
\frac{\Delta T}{T} = - \left\{
\begin{array}{ll}
\frac{\varepsilon}{\sqrt{1+4\left(\frac{\vartheta}{\vartheta_c}\right)^2}} & \mathrm{if}~~ \vartheta \leq \vartheta_* \\
&\\
\frac{\varepsilon}{2}\mathrm{e}^{-\frac{1}{2\vartheta^2_c}\left(\vartheta^2 + \vartheta^2_*\right)} & \mathrm{if}~~ \vartheta > \vartheta_* \\
\end{array}
\right. ,
\end{equation}
where $\vartheta$ is the angle from the center. $\varepsilon$ is
the amplitude parameter, and $\vartheta_c$ is the scale parameter.
$\vartheta_* = \sqrt{3}/2\vartheta_c$. By the Bayesian analysis,
the texture parameters were obtained
$\varepsilon=7.3^{+2.5}_{-3.6}\times10^{-5}$ and
$\vartheta_c=4.9^{+2.8}_{-2.4}{\rm deg}$ at $95\%$ confidence
\citep{cruz2007b}.

\begin{figure}
\includegraphics[width=80mm]{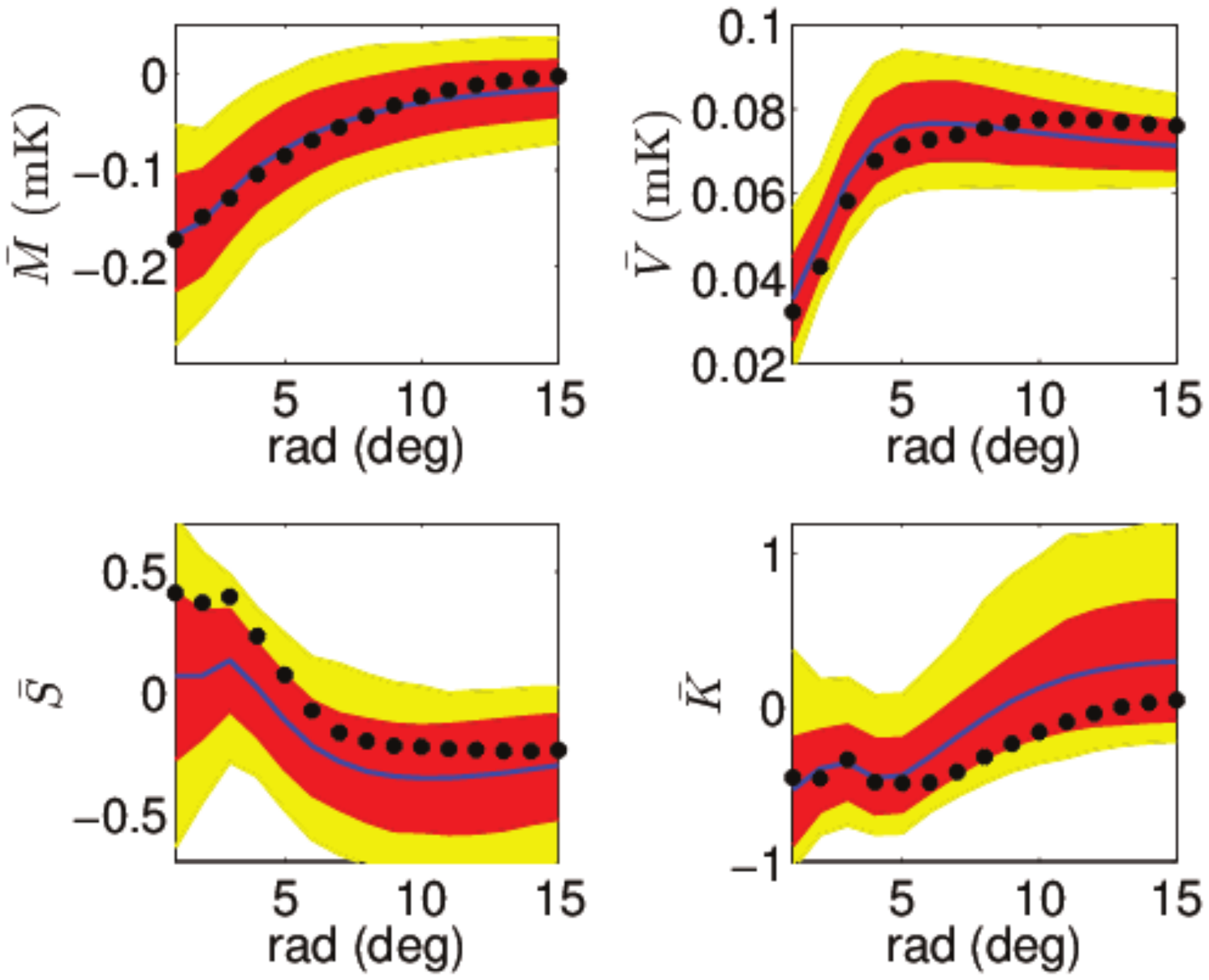}
    \caption{Four statistics for the spot at $(l=209^{\circ}, b=-57^{\circ})$. Confidence regions obtained from 500 Monte Carlo simulations are shown for 68 per cent (dark central region, red online) and
95 per cent (light outer region, yellow online) levels, as is the mean (solid blue line). Note that the cosmic texture is superimposed in each simulation. The observed statistics for WMAP ILC7 map are shown by the solid dots (black online).}
    \label{fig77}
\end{figure}

\begin{figure}
\includegraphics[width=80mm]{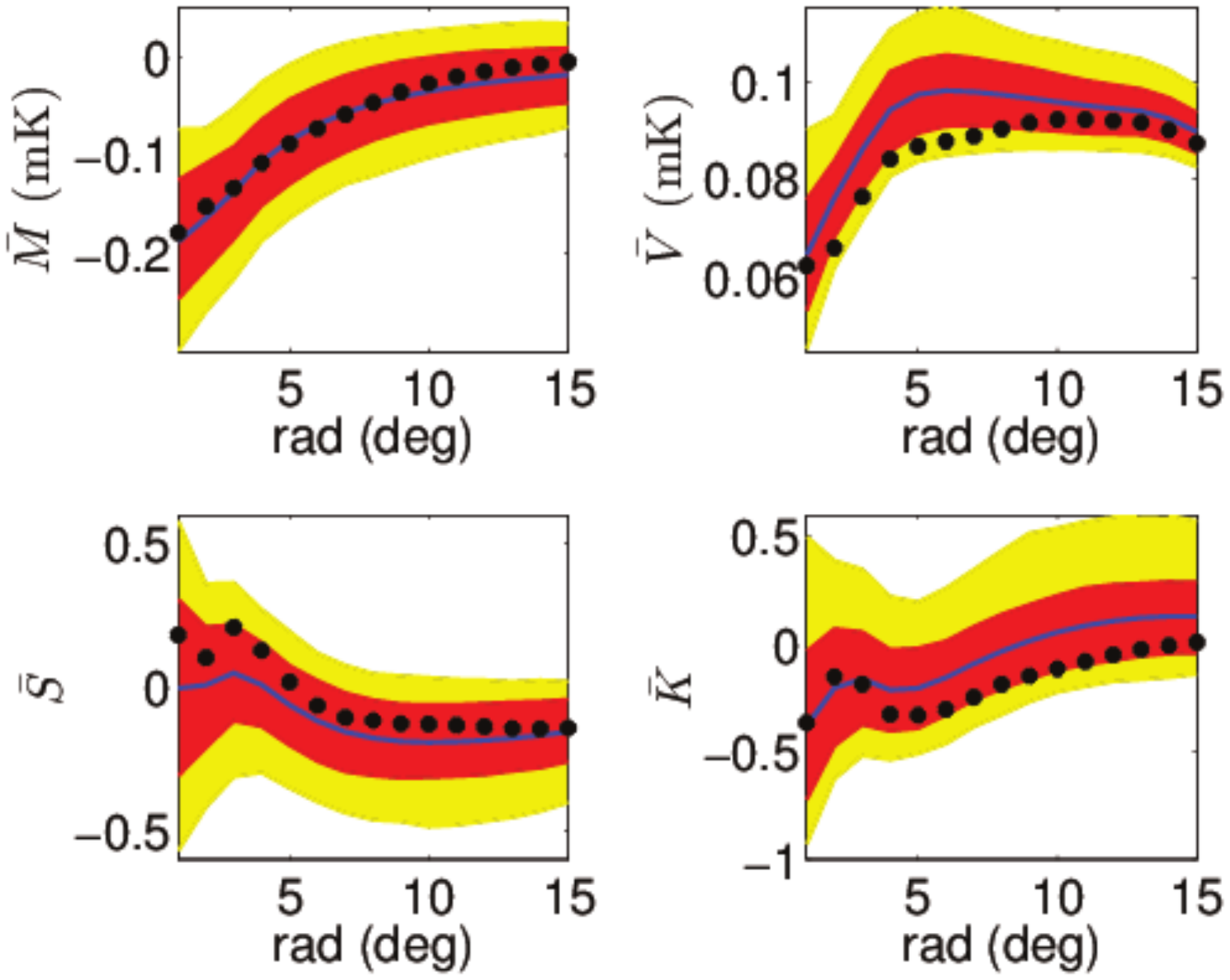}
    \caption{Same as Fig. \ref{fig77}, but ILC7 map is replaced by the masked VW7 maps.}
    \label{fig777}
\end{figure}

\begin{figure}
\includegraphics[width=80mm]{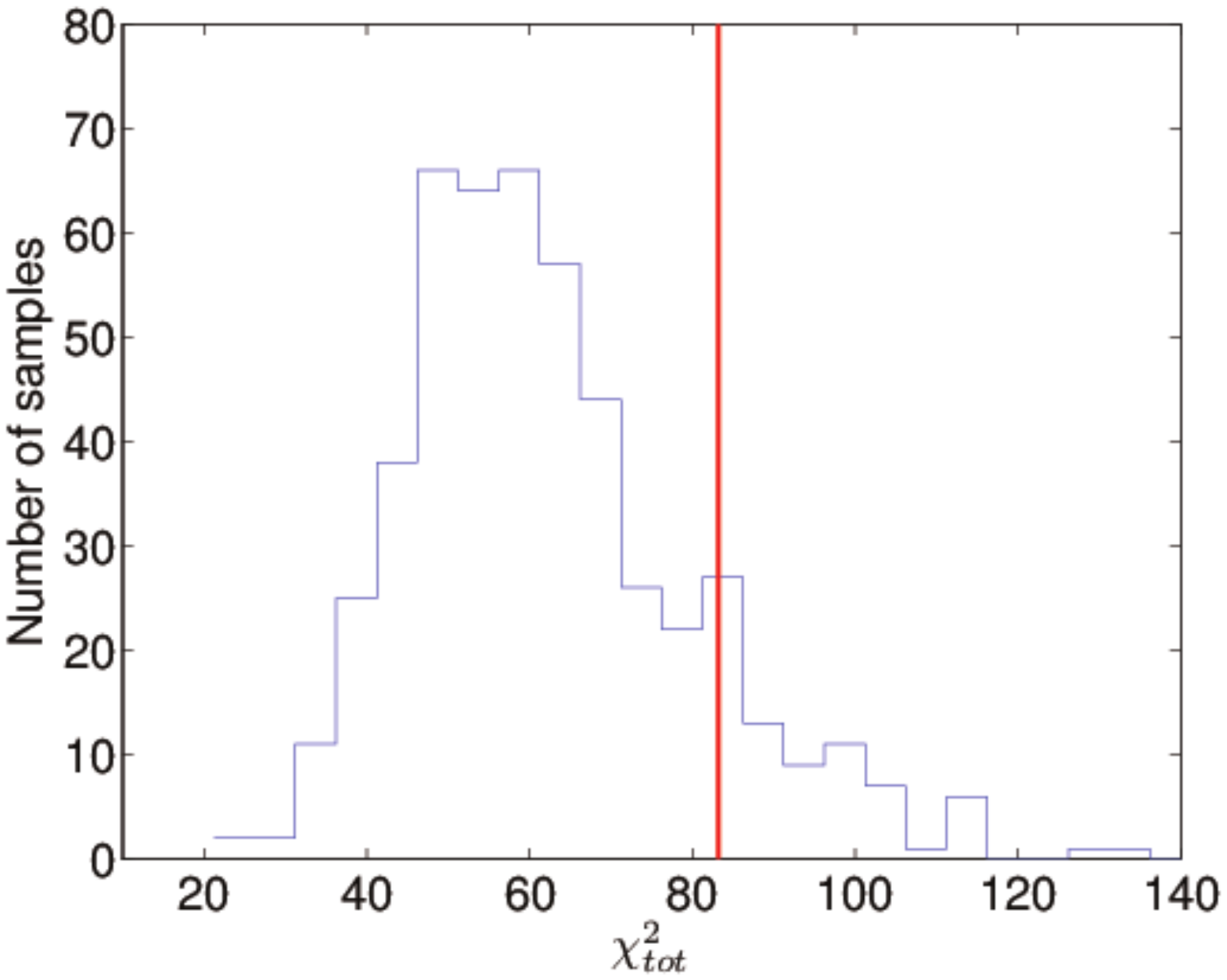}
    \caption{Histograms of the $\chi^2_{tot}$ statistic for the spots at $(l=209^{\circ}, b=-57^{\circ})$ obtained from 500 Monte Carlo simulations. Note that the cosmic texture is superimposed in each simulation. The observed statistic for WMAP ILC7 map is shown by the solid vertical line (red online).}
    \label{fig88}
\end{figure}

\begin{figure}
\includegraphics[width=80mm]{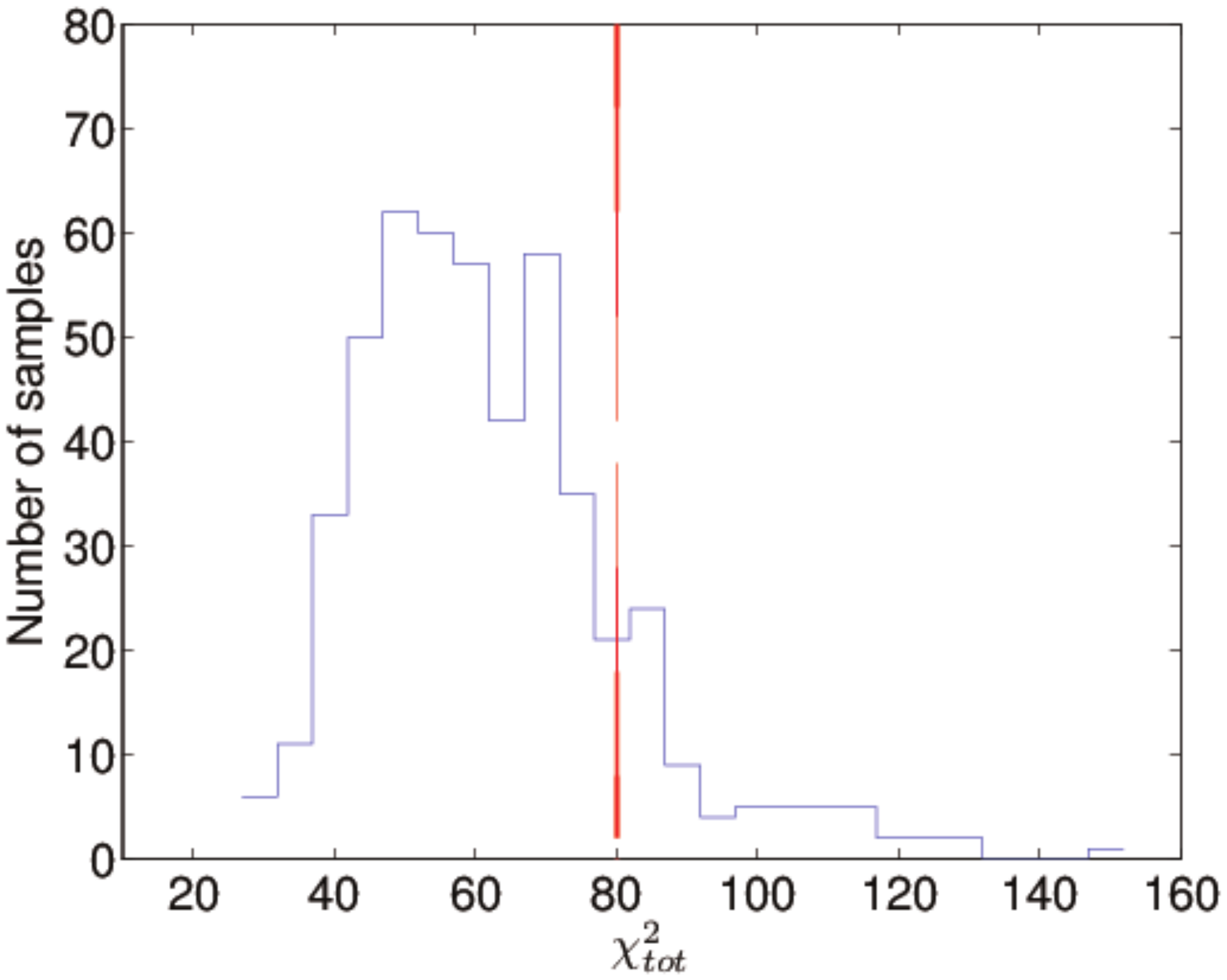}
    \caption{Same as Fig. \ref{fig88}, but ILC7 map is replaced by the masked VW7 maps.}
    \label{fig888}
\end{figure}

In our calculation, we adopt the best-fit texture parameters
$\varepsilon=7.3\times10^{-5}$ and $\vartheta_c=4.9^{\circ}$. Now, in order to taking the cosmic
texture into account, for each realization we superimpose the texture
morphology at the same position $(l=209^{\circ}, b=-57^{\circ})$,
and repeat the exactly same analyses. The results are also shown
in Figs. \ref{fig77}-\ref{fig888}. Interestingly enough, we find
that WMAP CS is exactly consistent with the simulations if the
cosmic texture is considered. For each statistic, the value of WMAP CS is equal to the mean value of
simulations in nearly 1-$\sigma$ confident level.

From Tables \ref{table1}-\ref{table3}, we find that
once the cosmic texture is considered in simulations, compared
with the pure Gaussian samples, every $\chi^2_{Y}$ value reduces,
especially for $Y=\bar{V}$ and $Y=\bar{S}$.
Note that the $\chi^2_{Y}$ value for $Y=\bar{M}$ has no significantly reduction,
which is caused by the following two facts: First, the cross-correlations
between $\bar{M}(R_i)$ and $\bar{M}(R_j)$ ($i\neq j$) are very strong;
Second, the values of $Y_0(R_i)-\bar{Y}(R_i)$ oscillate for different $R_i$.
For the $\chi_{tot}^2$ statistic of ILC7 CS, the significant level of the deviation from
simulations reduces to $90.0\%$, and that of VW7 CS becomes
$87.4\%$. So, we conclude that the cosmic texture can excellently
account for the excesses of $\bar{M}$, $\bar{V}$ and $\bar{S}$ of
WMAP CS, and the local analyses of WMAP CS strongly support the
cosmic texture explanation.

\begin{table}

 \centering

 \begin{minipage}{80mm}

  \caption{Same as Table \ref{table1}, but for the masked VW7 maps in Case 2 and Case 3.\label{table3}}

  \begin{tabular}{@{}llrrrrlrlr@{}}

  \hline
     & $\chi^2_{\bar{M}}$ & $\chi^2_{\bar{V}}$ & $\chi^2_{\bar{S}}$ & $\chi^2_{\bar{K}}$ & $\chi^2_{tot}$\\
\hline
Case 2  &  32.67    &  38.40   &  46.00   & 15.90    & 132.97    \\
Case 3  &  16.12    &  22.97   &  30.53   & 10.52    & 80.14     \\
\hline

\end{tabular}

\end{minipage}

\end{table}

\section{Cold spot and WMAP low multipoles}
If WMAP CS is a large-scale non-Gaussian structure, as we
have found in previous section, the
non-Gaussianity caused by CS should be encoded in the low
multipoles, rather than the high multipoles. In this section, we
shall confirm it by studying the effect of different multipoles on
the WMAP non-Gaussian signals.

Following \citep{vielva2004,cruz2005,cruz2006,zhang2010}, in this section we study the non-Gaussianity of WMAP data by using the wavelet transform, which can
emphasize or amplify some features of the CMB data at a particular scale. The SMHWs are defined as
  \begin{equation}
   \Psi(\theta;R)=A(R)(1+(\frac{y}{2})^2)^2
    (2-(\frac{y}{2})^2)e^{{-y^2}/{2R^2}},
  \end{equation}
where $y\equiv 2\tan(\theta/2)$ is the stereographic projection
variable, and $\theta\in [0,\pi)$ is the co-latitude. $R$ is the
scale, and $A$ is the constant for the normalization, which can be
written as
  \begin{equation}
   A(R)=\left[2\pi R^2\left(1+\frac{R^2}{2}+\frac{R^4}{4}\right)\right]^{-1/2}.
  \end{equation}

The continuous wavelet transform stereographically projected over the sphere with respect to $\Psi(\theta;R)$ is given by
   \begin{equation}
  T_{\rm w}({\hat{r}};R)=\int d\Omega' T({\hat{r}+\hat{r}'})\Psi(\theta';R), \label{smhw}
  \end{equation}
where ${\hat{r}}=(\theta,\phi)$ and ${\hat{r}'}=(\theta',\phi')$
are the stereographic projections to sphere of center of the spot
and the dummy location, respectively. In our analyses of this
section, the locations of centroids of spots are chosen to be
centers of pixels in $N_{\rm side}=32$ resolution. Following
\citep{zhang2010}, we define the occupancy fraction as follows to
account for the masked parts of the sky,
   \begin{equation}
  N_{\rm w}({\hat{r}};R)=\int d\Omega' M({\hat{r}+\hat{r}'})\Psi^2(\theta';R),
  \end{equation}
where $M(\hat{r})$ is KQ75y7 mask \citep{gold2011}. In order to
reduce the biases due to masking, we only include the results of
for which $N_{\rm w}({\hat{r}};R)>0.95$.

\begin{figure}
\includegraphics[width=80mm]{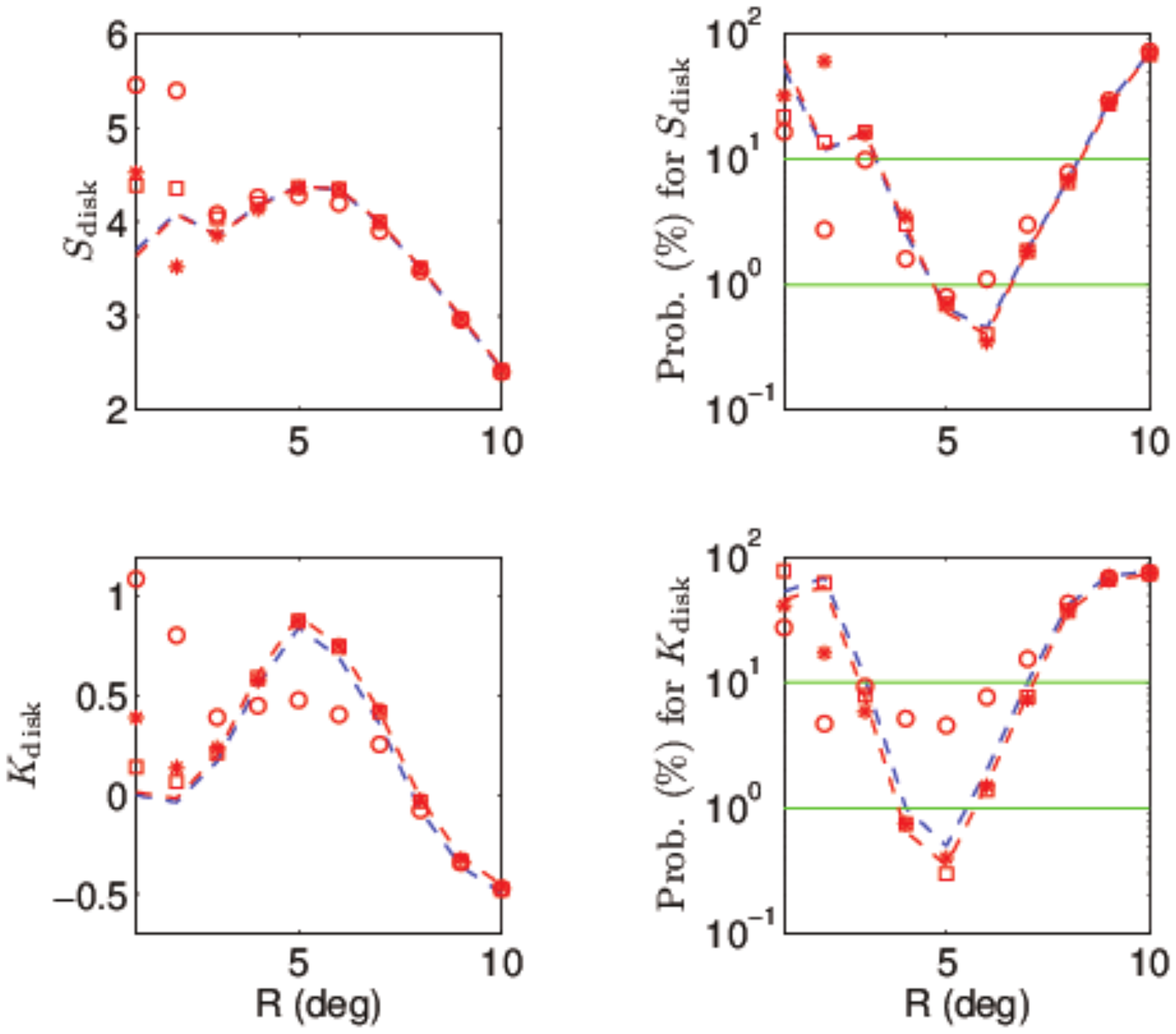}
    \caption{Values and probabilities of the statistics $S_{\rm disk}$ (upper) and $K_{\rm disk}$ (lower) for various cases.
 In each panel, the dark line (blue online) is for the results of VW7 map, and grey line (red online) is for ILC7 map.
 The \emph{circles} (red online) are for the ILC7 with $l_{\max}=20$, the \emph{crosses} (red online) are for that with $l_{\max}=40$,
 and the \emph{squares} (red online) are for that with $l_{\max}=100$.}
 \label{fig8}
\end{figure}

In our analyses of this section, we shall consider the VW7 and
ILC7 data. We degrade them to a lower resolution $N_{\rm
side}=128$, then apply the KQ75y7 mask. For each masked WMAP data,
we use the SMHW transform in Eq. (\ref{smhw}) to get the
corresponding map in wavelet domain $T_{\rm w}({\hat{r}};R)$. To
investigate the non-Gaussianity in different scales, for each map
we consider the cases with $R=1^{\circ}$, $2^{\circ}$,
$3^{\circ}$, $4^{\circ}$, $5^{\circ}$, $6^{\circ}$, $7^{\circ}$,
$8^{\circ}$, $9^{\circ}$, $10^{\circ}$. Similar to many authors
\citep{vielva2004,cruz2005,cruz2006,zhang2010}, we define the
statistics $S_{\rm disk}$ and $K_{\rm disk}$ as follows to study
the non-Gaussianity related to WMAP CS:
  \begin{eqnarray}
   S_{\rm disk}(R)&=& \frac{T_{\rm w}^{\rm coldest}(\hat{r};R)}{\sigma_{\rm w}(R)}, \\
   K_{\rm disk}(R)&=& \frac{1}{N_{\rm spots}} \frac{\sum_{i=1}^{N_{\rm spots}}T^4_{\rm w}(\hat{r}_i;R)}{\sigma^4_{\rm w}(R)}-3.
  \end{eqnarray}
Here $\sigma_{\rm w}(R)$ is the standard deviation of the
distribution of all spots in a given map, and $T_{\rm w}^{\rm
coldest}(\hat{r},R)$ is the \emph{coldest spot} in this distribution.
From the definitions, we find that $S_{\rm disk}(R)$ describes the
cold spot significance, and $K_{\rm disk}(R)$ is the kurtosis of
spots in a given map. In Fig. \ref{fig8} (left panels), we present
the values of $S_{\rm disk}(R)$ and $K_{\rm disk}(R)$ for
different scale parameter $R$. Both VW7 and ILC7 illustrate the
same results: the values of both $S_{\rm disk}(R)$ and $K_{\rm
disk}(R)$ maximize at $R\sim 5^{\circ}$. The results then are compared
with 2000 randomly generated Gaussian simulations, with the
exactly same methodology applied. So we can get the probabilities
of simulations, which have the larger $S_{\rm disk}(R)$ or $K_{\rm
disk}(R)$ than those of WMAP data. These probabilities for both
statistics are also shown in Fig. \ref{fig8} (right panel). So,
similar to other works \citep{vielva2004,cruz2005,zhang2010}, we
find that when $R=4^{\circ}-6^{\circ}$, WMAP data have the
deviations from the Gaussian simulations, i.e. the corresponding
probabilities for the statistics $S_{\rm disk}(R)$ and/or $K_{\rm
disk}(R)$ are smaller than $1\%$.

Now, let us study which multipoles account for the non-Gaussianity above. We consider the original ILC7 map $T(\hat{r})$, and expand it via spherical harmonic composition:
 \begin{equation}
 T(\hat{r})=\sum_{l,m}a_{lm} Y_{lm}(\hat{r}),
 \end{equation}
where $Y_{lm}$ are the spherical harmonics and $a_{lm}$ are the corresponding coefficients. Then the new map can be constructed as follows:
 \begin{equation}
 T'(\hat{r})=\sum_{l=2}^{l_{\max}} \sum_{m=-l}^{l}a_{lm} Y_{lm}(\hat{r}).
 \end{equation}
It is clear that this new map includes only the low multipoles
$l\in[2,~l_{\max}]$. Thus, we can repeat the processes above, but the ILC7 map $T(\hat{r})$ is replaced by $T'(\hat{r})$. In the
analyses, we choose three cases with $l_{\max}=20$, $l_{\max}=40$
and $l_{\max}=100$ to study the effect of different multipoles,
and show the results in Fig. \ref{fig8} with \emph{circles},
\emph{crosses} and \emph{squares}, respectively. We find that for
the statistic $S_{\rm disk}(R)$, if the lowest multipoles $l\le 20$
are considered, the values of the statistic and the corresponding
probabilities are quite close to those gotten in the map including
all the multipoles. These clearly show that the coldness of CS are
mainly encoded in these lowest multopole range, which is
consistent with the conclusion in \citep{naselsky2010}. While for
the statistic $K_{\rm disk}(R)$, we find the WMAP data are quite
normal for the case with $l_{\max}=20$, compared with the
Gaussian simulations. However, if $l_{\max}=40$ is considered, the
results for both statistics are very close to those in the map
with all multipoles. So we conclude that WMAP CS reflects directly
the peculiarities of the low multipoles $l\le 40$, which suggests
that CS should be a large-scale non-Gaussian structure, rather
than a combination of some small structures. This consists
with our conclusion in Section 3.

\section{Conclusions}
Since the discovery of the non-Gaussian Cold Spot in WMAP data, it has
attracted a great deal of attention, and many explanations have been
proposed. To distinguish them, in this paper we have studied the
local properties of WMAP CS at different scales by introducing the
local statistics including the mean temperature, variance,
skewness and kurtosis. Compared with the \emph{coldest spots} in random
Gaussian simulations, WMAP CS deviates from Gaussianity at $\sim
99\%$ significant level, and the non-Gaussianity of CS exists at
all the scales $1^{\circ}<R<15^{\circ}$. However, when compared
with the spots at the same position in the simulated Gaussian maps, we
found the significant excesses of local variance and skewness in the large
scales $R> 5^{\circ}$, rather than in the small scales.
Furthermore, we found that the non-Gaussianity caused by CS is
totally encoded in the WMAP low multipoles $l\le 40$. These all
imply that WMAP CS is a large-scale non-Gaussian structure,
rather than the combination of some small structures.

It was claimed by many authors that the cosmic texture with a
characteristic scale about $10^{\circ}$, rather than other
mechanisms, could provide the excellent explanation for WMAP CS.
By comparing with the random simulations including the similar
texture structure, we found this non-Gaussian structure could
excellently explain the excesses of the statistics. So our results
in this paper strongly support the cosmic texture explanation.

In the end of this paper, it is important to mention that the non-Gaussianity of WMAP CS has been confirmed by the new Planck observations \citep{planck2013} on the CMB temperature. In the near future, the polarization results of Planck mission will be released, which would play a crucial role to test the WMAP CS, as well as to reveal its physical origin \citep{cruz2007b}.

\section*{Acknowledgments}
We are very grateful to the anonymous referee for helpful remarks
and comments. We appreciate useful discussions with P. Naselsky, J. Kim, M. Hansen and A.M. Frejsel.
We acknowledge the use of the Legacy Archive for Microwave Background Data Analysis (LAMBDA).
Our data analysis made the use of HEALPix \citep{healpix} and GLESP \citep{glesp}.
This work is supported by NSFC No. 11173021, 11075141 and project
of Knowledge Innovation Program of Chinese Academy of Science.

\label{lastpage}

\end{document}